\documentclass[12pt]{article}
\usepackage{amsfonts}
\usepackage{amsmath}
\usepackage{amssymb}
\usepackage{amsthm}
\usepackage{accents}
\usepackage{youngtab}
\usepackage{braket}
\usepackage{graphicx}
\usepackage[svgnames]{xcolor}
\usepackage{ytableau,tikz}
\usepackage{here}
\usepackage{comment}
\usepackage{simplewick}
\usepackage{tikz}
\usepackage{feynmf}

\newcommand{\ba}{\begin{eqnarray}}
\newcommand{\ea}{\end{eqnarray}}
\newcommand{\nn}{\nonumber}

\newcommand{\cN}{\mathcal{N}}
\newcommand{\cW}{\mathcal{W}}

\newcommand{\cY}{\mathcal{Y}}
\newcommand{\lt}{\left}
\newcommand{\rt}{\right}

\newcommand{\gammah}{\hat{\gamma}}

 \DeclareMathOperator*{\Res}{Res}

\makeatletter
\newcommand{\douwidehat}[2]{%
  \sbox0{$\m@th#1\widehat{\hphantom{#2}}$}%
  \sbox2{$\m@th#1x$}
  \sbox4{$\m@th#1#2$}
  \dimen0=\ht0
  \advance\dimen0 -.8\ht2
  \dimen2=\dp4
  \rlap{%
    \raisebox{\dimexpr\dimen0-\dimen2}{%
      \scalebox{1}[-1]{\box0}%
    }%
  }%
  {#2}%
}
\makeatother

\makeatletter
\@addtoreset{equation}{section}

\makeatother

\usepackage[colorlinks=true,linkcolor=black,citecolor=black,urlcolor=black]{hyperref}

\begin{document}

\begin{titlepage}
\vspace*{-2cm}
	\begin{flushright}
		UT-17-34
	\end{flushright}
	
	\vskip 12mm
	
	\begin{center}
		{\Large\bf An Elliptic Vertex of Awata-Feigin-Shiraishi type for M-strings}
	\end{center}
	\vfill
		\begin{center}
			{\Large Rui-Dong Zhu}
			\\[.6cm]
			{\em {} Department of Physics, The University of Tokyo}\\
			{\em Bunkyo-ku, Tokyo, Japan}
			\\[.4cm]
		\end{center}
	\vfill
	\begin{abstract}
		We write down a vertical representation for the elliptic Ding-Iohara-Miki algebra, and construct an elliptic version of the refined topological vertex of Awata, Feigin and Shiraishi . We show explicitly that this vertex reproduces the elliptic genus of M-strings, and that it is an intertwiner of the algebra.
	\end{abstract}
	\vfill
\end{titlepage}

\section{Introduction}

The AGT relation \cite{AGT,Wyllard09} is an intriguing string duality connecting 4d $\cN=2$ gauge theories and 2d CFTs. It was inspired by the class ${\cal S}$ construction \cite{classS} of $\cN=2$ gauge theories by compactifying certain 6d $\cN=(2,0)$ SCFT on a punctured Riemann surface. There is a similar proposal \cite{5dAGT} for 5d $\cN=1$ gauge theories, whose dual is proposed to be $q$-deformed $\cW$-algebras. The class ${\cal S}$ picture clearly does not explain the origin of this $q$-deformed $\cW$-symmetry, and that the 5d AGT relation reduces to the original AGT proposal in the $q\rightarrow 1$ limit suggests that another picture might exist for this kind of relations, which can be converted to the class S picture through a series of string dualities. Interestingly, in 2011, an alternative but equivalent realization of the refined topological vertex to the original Iqbal-Kozcaz-Vafa (IKV) version \cite{IKV} was reported by Awata, Feigin, and Shiraishi (AFS) \cite{AFS}, to be identified as the intertwiner\footnote{The IKV refined topological vertex is the tensor element of this intertwiner under some proper basis.} in a so-called Ding-Iohara-Miki (DIM) algebra \cite{DI, Miki}. This underlying algebra is established to be a $q$-deformed $\cW_{1+\infty}$ algebra \cite{Miki}, and in \cite{Prochazka} (see also \cite{FMNZ}), it is conjectured to be the same $\cW_\infty[\mu]$ algebra used in the higher-spin AdS$_3$/CFT$_2$ holography \cite{AdS/minimal} up to a u(1)-extension. In \cite{AFS}, the $(p,q)$-brane web \cite{AHK} used to construct 5d gauge theories, which is dual to the toric diagram of the toric Calabi-Yau manifold in the topological string approach \cite{Leung-Vafa}, is interpreted as a web of representations of the DIM algebra. This suggests a potential connection between the toric diagram and the origin of the $\cW$-symmetry of gauge theories. Indeed, a linear $\cW_{1+\infty}$ symmetry is discovered in the B-model on the mirror Calabi-Yau \cite{ADKMV} in the unrefined limit. There is another type of $q$-deformed $\cW$-algebra associated to the quiver of 5d gauge theories discovered in \cite{Kimura-Pestun}. Let us denote this $\cW$-algebra associated to the quiver $\Gamma$ as $\cW_{q,t}(\Gamma)$ and the $\cW$-lagebra associated to the gauge group $\mathfrak{g}$ in the original AGT proposal as $\cW_{q,t}(\mathfrak{g})$ (these $q$-deformed $\cW$-algebras associated to Lie algebra $\mathfrak{g}$, $\cW_{q,t}(\mathfrak{g})$, are the same as those defined in \cite{Frenkel-Reshetikhin}). $\cW_{q,t}(\Gamma)$ transforms to $\cW_{q,t}(\mathfrak{g})$ under the fiber-base duality \cite{fiber-base}, which interchanges D5-branes and NS5-branes in terms of the brane web. Interestingly, the fiber-base duality can be found in the ${\rm SL}(2,\mathbb{Z})$ automorphism \cite{Miki} of the DIM algebra, as explained in \cite{BFHMZ} that both $\cW_{q,t}(\mathfrak{g})$ and $\cW_{q,t}(\Gamma)$ are embedded respectively in the vertical and horizontal representation of the DIM algebra. More generally, truncation of the $(p,q)$-brane web at different angles gives rise to various $\cW$-algebras as discussed in \cite{VOAcorner}. It is then natural to expect a detailed investigation on the DIM algebra will reveal the nature of the AGT relation, even though neither the field-theory realization of $q$-$\cW$ algebras nor the worldsheet formulation of the refined topological string is clear at the current stage. 

It was soon discovered that the partition function of 6d $\cN=(1,0)$ theory realized by compactified $(p,q)$-brane web along the NS5-brane direction\cite{M-strings} also has interpretation as a conformal block in the elliptic Virasoro algebra \cite{ellip-Virasoro,IKY}. The 6d partition function is computed with the refined topological vertex on a compactified toric Calabi-Yau \cite{HIV, M-strings,ellip-Virasoro}, and the system is argued to be dual to the system of M2 branes suspended between parallel M5 branes, or the M-strings (on $A$-type ALE space \cite{M-orbifold}). An alternative derivation of the 5d and 6d AGT relation from the M-string was provided in \cite{Tan-13, Tan-16}. On the other hand, a natural elliptic extension of the DIM algebra was constructed in a rather different context \cite{Saito}, and we refer to this algebra as the elliptic DIM algebra in this article. As already discussed in \cite{ellip-Virasoro}, the elliptic Virasoro algebra can be embedded into the coproduct of two horizontal representations of the elliptic DIM algebra constructed in \cite{Saito}, and thus it is natural to expect that one can formulate the M-strings computation in a similar way to \cite{AFS} with a web of representations of the elliptic DIM algebra. 

In this paper, we first write down a representation of the elliptic DIM algebra in analogy to the vertical representation of the usual DIM algebra, and then we construct an elliptic version of the refined topological vertex. As the main claim of this paper, we prove the AFS property, which states that the elliptic AFS vertex is indeed an intertwiner of the elliptic DIM algebra (the algebra acts in the adjoint way to the vertex). At the end, 
we conclude the article by discussing the application of these vertices to the elliptic AGT relation. 

The following figure sketches another point of this paper: the elliptic genus of M-strings (instanton partition function of 6d theories) can be produced with the elliptic vertex introduced in this paper. 
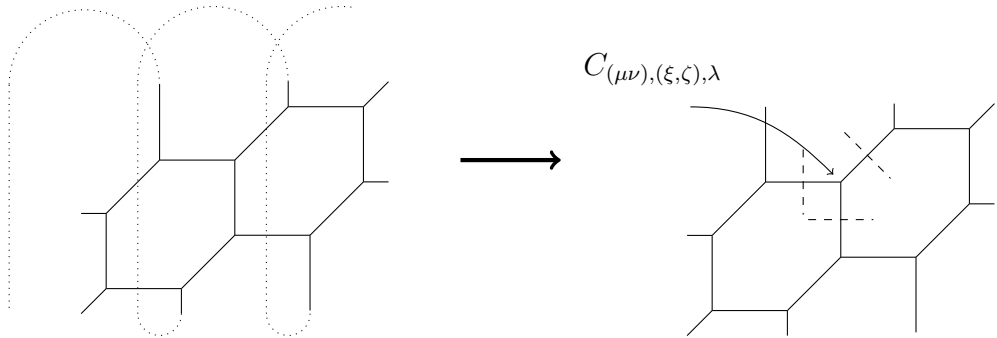
\begin{center}
      \begin{tikzpicture}[scale=1/3]
      \draw (0,0)--(3,0);
      \draw (0,0)--(0,3);
      \draw (0,0)--(-2.13,-2.13);
      \draw (3,0)--(5.13,2.13);
      \draw (3,0)--(3,-3);
      \draw (5.13,2.13)--(5.13,3.13);
      \draw (5.13,2.13)--(8.13,2.13);
      \draw (3,-3)--(0.87,-5.13);
      \draw (0.87,-5.13)--(-2.13,-5.13);
      \draw (-2.13,-2.13)--(-2.13,-5.13);
      \draw (3,-3)--(6,-3);
      \draw (6,-3)--(8.13,-0.87);
      \draw (8.13,-0.87)--(8.13,2.13);
      \draw (8.13,2.13)--(9.13,3.13);
      \draw (8.13,-0.87)--(9.13,-0.87);
      \draw (6,-3)--(6,-6);
      \draw (0.87,-5.13)--(0.87,-6.13);
      \draw (-2.13,-5.13)--(-3.13,-6.13);
      \draw (-2.13,-2.13)--(-3.13,-2.13);
      \draw[dotted] (5.13,3+0.13) arc (0:180:3);
      \draw[dotted] (0.87,-6.13) arc (0:-180:0.87);
      \draw[dotted] (5.13-6,-6.13)--(5.13-6,3.13);
      \draw[dotted] (0,3) arc (0:180:3);
      \draw[dotted] (-6,3)--(-6,-6);
       \draw[dotted] (0.87+5.13,-6.13) arc (0:-180:0.87);
      \draw[dotted] (5.13-6+5.13,-6.13)--(5.13+5.13-6,3.13);
      \draw[dotted] (5.13+5.13-6,3+0.13) arc (180:80:3);
      \draw[ultra thick,->] (12,0)--(16,0);
      \end{tikzpicture}
 \begin{tikzpicture}[scale=1/3]
      \node at (-4.5,4.5) {$C_{(\mu\nu),(\xi,\zeta),\lambda}$};
      \draw (0,0)--(3,0);
      \draw (0,0)--(0,3);
      \draw (0,0)--(-2.13,-2.13);
      \draw (3,0)--(5.13,2.13);
      \draw (3,0)--(3,-3);
      \draw (5.13,2.13)--(5.13,3.13);
      \draw (5.13,2.13)--(8.13,2.13);
      \draw (3,-3)--(0.87,-5.13);
      \draw (0.87,-5.13)--(-2.13,-5.13);
      \draw (-2.13,-2.13)--(-2.13,-5.13);
      \draw (3,-3)--(6,-3);
      \draw (6,-3)--(8.13,-0.87);
      \draw (8.13,-0.87)--(8.13,2.13);
      \draw (8.13,2.13)--(9.13,3.13);
      \draw (8.13,-0.87)--(9.13,-0.87);
      \draw (6,-3)--(6,-6);
      \draw (0.87,-5.13)--(0.87,-6.13);
      \draw (-2.13,-5.13)--(-3.13,-6.13);
      \draw (-2.13,-2.13)--(-3.13,-2.13);
      \draw[dashed] (1.5,1.3)--(1.5,-1.3);
      \draw[dashed] (1.7,-1.5)--(4.3,-1.5);
      \draw[dashed] (3.145,1.985)--(4.985,0.145);
      \draw[->] (-3,3) to [out=0, in=135] (2.7,0.3);
      \end{tikzpicture}
           \end{center}
           
\paragraph{Notation related to Young diagrams} For a given Young diagram $\lambda$, we decompose it into a series of non-negative integer numbers $\{\lambda_i|i\in\mathbb{Z_+}\}$ satisfying, $\lambda_1\geq \lambda_2\geq \lambda_3\geq \dots$. These numbers correspond to the number of boxes in each column of the Young diagram. We denote $\lambda^t$ as the transpoed Young diagram of $\lambda$, and $(i,j)\in\lambda$ stands for the box in the $i$-th row and $j$-th column in $\lambda$. All notations used in this article are listed in the following table. 
\begin{center}
\begin{tabular}{|p{7cm}|c|p{7cm}|}
\hline
terminology & notation & meaning\\
\hline
arm length & $a(i,j)$ & $a(i,j)=\lambda^t_j-i$\\
\hline
leg length & $\ell(i,j)$ & $\ell(i,j)=\lambda_i-j$\\
\hline
coordinate of a box $x=(i,j)\in \lambda$ & $\chi_x$ & $\chi_x=vq_1^{i-1}q_2^{j-1}$ with $v$ the highest-weight parameter of $\lambda$\\
\hline
set of all boxes that can be added to $\lambda$ & $A(\lambda)$ & see Figure \ref{Young-diagram}\\
\hline 
set of all boxes that can be removed from $\lambda$ & $R(\lambda)$ & see Figure \ref{Young-diagram}\\
\hline 
size of $\lambda$ & $|\lambda|$ & $|\lambda|=\sum_i\lambda_i$\\
\hline
\end{tabular}
\end{center}

\begin{figure}
\begin{center}
\begin{tikzpicture}[scale=0.5]
\draw (0,0)--(5,0);
\draw (5,0)--(5,-2);
\draw (5,-2)--(3,-2);
\draw (3,-2)--(3,-3);
\draw (3,-3)--(1,-3);
\draw (1,-3)--(1,-5);
\draw (1,-5)--(0,-5);
\draw (0,-5)--(0,0);
\draw [fill=gray] (0,-5) rectangle (1,-6);
\draw [fill=gray] (1,-3) rectangle (2,-4);
\draw [fill=gray] (3,-2) rectangle (4,-3);
\draw [fill=gray] (5,0) rectangle (6,-1);
\draw [dotted] (0,-4)--(1,-4);
\draw [dotted] (2,-2) rectangle (3,-3);
\draw [dotted] (4,-1) rectangle (5,-2);
\end{tikzpicture}
\caption{An example of Young diagram $\lambda$ and boxes in $A(\lambda)$ (shaded) and in $R(\lambda)$ (dotted). (Boxes in $A(\lambda)$ do not belong to $\lambda$, and $R(\lambda)\subset \lambda$. We always have $|A(\lambda)|-|R(\lambda)|=1$. )}
\label{Young-diagram}
\end{center}
\end{figure}
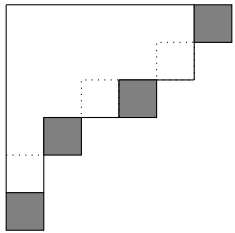

The Frobenius coordinate $(\alpha_1,\alpha_2,\dots,\alpha_s|\beta_1,\beta_2,\dots,\beta_s)$ of a Young diagram $\lambda$ is defined by 
\ba
\alpha_i=a(i,i)+\frac{1}{2},\quad \beta_i=\ell(i,i)+\frac{1}{2},\quad {\rm for}\ i=1,2,\dots,s.\label{Frobenius}
\ea
$s$ represents the number of boxes in the diagonal direction of $\lambda$. Note that $\sum_i(\alpha_i+\beta_i)=|\lambda|$ and exchanging $\alpha_i$ and $\beta_i$ transforms $\lambda$ to $\lambda^t$.

\section{Elliptic Genus of M-strings}

The elliptic genus of M-strings \cite{M-strings} can be computed from the refined topological string on a Calabi-Yau geometry specified by a toric diagram with its top and bottom external legs identified together. It can also be interpreted as the instanton partition function of the corresponding 6d $\cN=(1,0)$ theory on $\mathbb{R}^4\times T^2$ with omega background twisting. This partition function is an elliptic version of the Nekrasov partition function, which is a natural consequence from the fact that the refined topological vertex on toric Calabi-Yau is dual to 5d $\cN=1$ theory constructed from $(p,q)$-brane web, and the compactification (i.e. identification of external legs) of this brane web along the NS5 direction lifts the theory to 6d. We give a brief review on it in this section, by rederiving it on the compactified toric Calabi-Yau, with the Awata-Feigin-Shiraishi (AFS) version of the refined topological vertex \cite{AFS} (see also \cite{BFHMZ} for a review and generalization). The derivation for the partition function of a strip with arbitrary even number of horizontal legs was originally done in \cite{ellip-Virasoro}. The building block for a compactified brane web is 
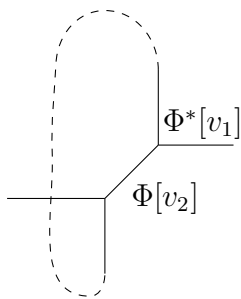
\begin{center}
\begin{tikzpicture}
\draw(0.7,0)-- (2,0);
\draw (2,0)--(2.71,0.71);
\draw (2,0)--(2,-1);
\draw (2.71,0.71)--(3.71,0.71);
\draw (2.71,0.71)--(2.71,1.71);
\draw[dashed] (2.71,1.71) to [out=90,in=0] (2,2.5) to [out=180,in=87] (1.35,1.8) to [out=-93,in=92] (1.3,0.6) to [out=-88,in=180] (1.8,-1.3) to [out=0,in=-90] (2,-1);
\node at (3.3,1) {$\Phi^\ast[v_1]$};
\node at (2.8,0) {$\Phi[v_2]$};
\end{tikzpicture}
\end{center}
We note that by replacing $\Phi^\ast$ and $\Phi$ with the generalized AFS vertices introduced in \cite{BFHMZ}, it is possible to build the compactified brane web with arbitrary rank of gauge group. What we need to do is to take the trace over the Fock space of the horizontal representation of the DIM algebra. There are various of ways to compute the trace in the Fock space, and one of them is reviewed in Appendix \ref{a-trace}. The trace of $:\Phi[v_2]\Phi^\ast[v_1]$ can be evaluated to\footnote{We used a convenient notation in this equation that whenever $x$, or $x'$ (resp. $y$ or $y'$) appears, it is summed over all boxes in $x,x'\in\lambda$ (resp. $y,y'\in\mu$). This convenient notation is used throughout this section. } 
\ba
&&\frac{\prod_{n>0}1-p^{n}}{t^\ast(\lambda,v_1)t(\mu,v_2)}{\rm tr}_\tau:\Phi^\ast_\lambda[v_1]\Phi_\mu[v_2]:\nn\\
&&=\exp\lt(-\sum_{n>0}\frac{1}{n}\frac{p^n}{1-p^n}\frac{1}{(1-t^n)(1-q^{-n})}\lt(1+q_3^n-\gamma^n\lt(\frac{v_1}{v_2}\rt)^n-\gamma^n\lt(\frac{v_2}{v_1}\rt)^n\rt.\rt.\nn\\
&&-(1-t^n)(1-q^{-n})\lt(\lt(\frac{\chi_y}{v_2}\rt)^n-\gamma^{-n}\lt(\frac{\chi_y}{v_1}\rt)^n-\gamma^{-n}\lt(\frac{\chi_x}{v_2}\rt)^n+q_3^{-n}\lt(\frac{\chi_x}{v_1}\rt)^n\rt)\nn\\
&&-(1-t^n)(1-q^{-n})\lt(\lt(\frac{v_1}{\chi_x}\rt)^n-\gamma^n\lt(\frac{v_1}{\chi_y}\rt)^n-\gamma^n\lt(\frac{v_2}{\chi_x}\rt)^n+q_3^n\lt(\frac{v_2}{\chi_y}\rt)^n\rt)\nn\\
&&\lt.+(1-t^n)^2(1-q^{-n})^2\lt(q_3^{-n}\lt(\frac{\chi_x}{\chi_{x'}}\rt)^n-\gamma^{-n}\lt(\frac{\chi_y}{\chi_x}\rt)^n-\gamma^{-n}\lt(\frac{\chi_x}{\chi_y}\rt)^n-\lt(\frac{\chi_y}{\chi_{y'}}\rt)^n\rt)\rt),\label{eq-1}
\ea
where $q_3:=t/q=\gamma^2$, $p=e^{-\tau}$ is the modulus of the torus, and the AFS vertices are given by 
\ba
\Phi^{(n)}_\lambda[u,v]=t_n(\lambda,u,v):\Phi_\emptyset(v)\prod_{x\in \lambda}\eta(\chi_x):,\\
\Phi^{\ast(n)}_\lambda[u,v]=t^\ast_n(\lambda,u,v):\Phi^\ast_\emptyset (v)\prod_{x\in \lambda}\xi(\chi_x),
\ea
with the normalization factors 
\ba
t_n(\lambda,u,v)=(-uv)^{|\lambda|}\prod_{x\in\lambda}(\gamma/\chi_x)^{n+1},\\
t^\ast_n(\lambda,u,v)=(\gamma u)^{-|\lambda|}\prod_{x\in \lambda}(\chi_x/\gamma)^n,
\ea
and the vertex operators, 
\ba
V_\pm (z)=\exp\lt(\pm \sum_{n\geq 1}\frac{1}{n}\frac{z^{\mp n}}{1-q^{\mp n}}a_{\pm n}\rt),\\
\Phi_\emptyset(v)=V_-(v)V_+(v),\\
\Phi^\ast_\emptyset(v)=V_-^{-1}(\gamma v)V_+^{-1}(v/\gamma),
\ea
\ba
\eta(z)=\exp\lt(\sum_{n\geq 1}\frac{1-t^{-n}}{n}z^na_{-n}\rt)\exp\lt(-\sum_{n\geq 1}\frac{1-t^n}{n}z^{-n}a_n\rt),\\
\xi(z)=\exp\lt(-\sum_{n\geq 1}\frac{1-t^{-n}}{n}\gamma^nz^na_{-n}\rt)\exp\lt(\sum_{n\geq 1}\frac{1-t^n}{n}\gamma^nz^{-n}a_n\rt),
\ea
expressed in $q$-bosons satisfying 
\ba
[a_n,a_{-m}]=n\frac{1-q^{|n|}}{1-t^{|n|}}\delta_{n,m}.
\ea
$q$ and $t$ are two $\Omega$-background parameters, and $\gamma$ is a short notation for $\gamma=(t/q)^{1/2}$. $\chi_x$ is a coordinate-like complex number assigned to each box $x$ in the Young diagram $\lambda$ associated to the Coulomb branch parameter\footnote{To be more precise, it is the position parameter for the associated D5 brane in the brane web. The ratio of two $v$'s parameterizes the Coulomb branch. } $v$. When $x$ is in the $i$-th row and $j$-th column, $\chi_x=vq^{i-1}t^{-j+1}$. We note that $u$-dependence of the AFS vertex was not explicitly written in the l.h.s. of equation (\ref{eq-1}), as it only appears in the overall normalization factor $t$ and $t^\ast$ and cancels out on the r.h.s. Refer to \cite{AFS, BFHMZ} for more details about the AFS vertex. 

Equation (\ref{eq-1}) can be divided into four parts. First we have an overall factor which does not depend on $\lambda$ or $\mu$, i.e. terms in the first line of equation (\ref{eq-1}) combined with a similar factor from the contraction of $\Phi^\ast_\lambda$ and $\Phi_\mu$, 
\ba
{\cal G}^{ellip}(\gamma v_1,v_2;q,t,p):=\exp\lt[\sum_{n=1}^\infty\frac{1}{n}\frac{1}{(1-q^{-n})(1-t^n)}\lt((\gamma v_1/v_2)^n\rt.\rt.\nn\\
\lt.\lt.-\frac{p^n}{1-p^n}\lt(1+q_3^n-\gamma^n\lt(\frac{v_1}{v_2}\rt)^n-\gamma^n\lt(\frac{v_2}{v_1}\rt)^n\rt)\rt)\rt].\label{G-ellip}
\ea
This is an elliptic version of the original overall factor ${\cal G}(\gamma v_1,v_2;q,t):=\exp\lt(\sum_{n=1}^\infty\frac{1}{n}\frac{(\gamma v_1/v_2)^n}{(1-q^{-n})(1-t^n)}\rt)$ with an additional new prefactor 
\ba
{\cal Z}^{ellip}:=\exp\lt(-\sum_{n=1}^\infty\frac{1}{n}\frac{1+q_3^n}{(1-q^{-n})(1-t^n)(1-p^n)}\rt).
\ea
The second part depends on both $\lambda$ and $\mu$, and it is an elliptic version of the bifundamental contribution. Up to the original bifundamental contribution coming from the contraction of $\Phi^\ast$ and $\Phi$, 
\ba
Z_{bfd}(v_1,\lambda,v_2,\mu|\gamma^{-1})=\prod_{x\in\lambda,y\in\mu}\lt(1-\frac{q\gamma \chi_x}{tv_2}\rt)\lt(1-\frac{\gamma v_1}{\chi_y}\rt)\frac{\lt(1-\frac{\gamma q\chi_x}{\chi_y}\rt)\lt(1-\frac{t^{-1}\gamma\chi_x}{\chi_y}\rt)}{\lt(1-\frac{\gamma\chi_x}{\chi_y}\rt)\lt(1-\frac{qt^{-1}\gamma\chi_x}{\chi_y}\rt)},
\ea
and we have in addition, 
\ba
\exp\lt(-\sum_{n>0}\frac{1}{n}\frac{p^n}{1-p^n}\lt(\gamma^{-n}\lt(\frac{\chi_y}{v_1}\rt)^n+\gamma^{-n}\lt(\frac{\chi_x}{v_2}\rt)^n+\gamma^n\lt(\frac{v_1}{\chi_y}\rt)^n+\gamma^n\lt(\frac{v_2}{\chi_x}\rt)^n\rt.\rt.\nn\\
\lt.\lt.-(1-t^{-n})(1-q^n)\lt(\gamma^{n}\lt(\frac{\chi_y}{\chi_x}\rt)^n+\gamma^{n}\lt(\frac{\chi_x}{\chi_y}\rt)^n\rt)\rt)\rt)\nn\\
=\lt(p\gamma^{-1}\frac{\chi_y}{v_1};p\rt)_\infty\lt(p\gamma^{-1}\frac{\chi_x}{v_2};p\rt)_\infty\lt(p\gamma\frac{v_1}{\chi_y};p\rt)_\infty\lt(p\gamma\frac{v_2}{\chi_x};p\rt)_\infty\nn\\
\times\frac{\lt(pt^{-1}\gamma\frac{\chi_y}{\chi_x};p\rt)_\infty\lt(pq\gamma\frac{\chi_y}{\chi_x};p\rt)_\infty}{\lt(p\gamma\frac{\chi_y}{\chi_x};p\rt)_\infty\lt(pq_3^{-1}\gamma\frac{\chi_y}{\chi_x};p\rt)_\infty}
\frac{\lt(pt^{-1}\gamma\frac{\chi_x}{\chi_y};p\rt)_\infty\lt(pq\gamma\frac{\chi_x}{\chi_y};p\rt)_\infty}{\lt(p\gamma\frac{\chi_x}{\chi_y};p\rt)_\infty\lt(pq_3^{-1}\gamma\frac{\chi_x}{\chi_y};p\rt)_\infty},
\ea
which combine together into 
\ba
Z^{ellip}_{bfd}(v_1,\lambda;v_2,\mu|\gamma^{-1})=\prod_{y\in\mu}\theta_p(\gamma v_1/\chi_y)\prod_{x\in\lambda}\theta_p(\gamma^{-1} \chi_x/v_2)\prod_{x\in\lambda,y\in\mu}\frac{\theta_p(t^{-1}\gamma\chi_x/\chi_y)\theta_p(q\gamma\chi_x/\chi_y)}{\theta_p(\gamma\chi_x/\chi_y)\theta_p(q_3^{-1}\gamma\chi_x/\chi_y)},\label{ellip-bifund}
\ea
where the $\theta$-function $\theta_p(z)$ is defined by 
\begin{equation}
\theta_p(z) = \lt( z; p \rt)_\infty \, \lt( p z^{-1}; p \rt)_\infty = 
\prod_{k \geq 0} \lt( 1 - p^k     z      \rt) \,  
\prod_{k \geq 0} \lt( 1 - p^{k+1} z^{-1} \rt).
\end{equation}
This is precisely the elliptic Nekrasov factor, and we review its several equivalent expressions in Appendix \ref{Nekra}.

The remaining two parts correct the vector multiplet contributions of $\lambda$ and $\mu$ respectively. The factor associated to $\lambda$ reads 
\ba
\exp\lt[\sum_{n>0}\frac{1}{n}\frac{p^n}{1-p^n}\lt(q_3^{-n}\lt(\frac{\chi_x}{v_1}\rt)^n+\lt(\frac{v_1}{\chi_x}\rt)^n-(1-t^n)(1-q^{-n})q_3^{-n}\lt(\frac{\chi_x}{\chi_{x'}}\rt)^n\rt)\rt],\label{lambda-fac}
\ea
and that for $\mu$ is given by 
\ba
\exp\lt[\sum_{n>0}\frac{1}{n}\frac{p^n}{1-p^n}\lt(\lt(\frac{\chi_y}{v_2}\rt)^n+q_3^n\lt(\frac{v_2}{\chi_y}\rt)^n-(1-t^n)(1-q^{-n})\lt(\frac{\chi_y}{\chi_{y'}}\rt)^n\rt)\rt].\label{mu-fac}
\ea
Note that the vector multiplet contribution comes from the inner product of $\Phi^\ast$ and $\Phi$, it is corrected by both a (\ref{lambda-fac})-like factor and also a (\ref{mu-fac})-like factor. Taking into account of the $\gamma^{-1}$ shift in (\ref{ellip-bifund}), we find that 
\ba
Z_{vec}^{ellip}(v,\lambda)=Z_{bfd}(v,\lambda;v,\lambda|1)^{-1}.\label{vec-ellip}
\ea

Therefore, we see that the partition function of the 6d theory on $T^2$ can also be built with building blocks of $Z_{bfd}^{ellip}$ and $Z_{vec}^{ellip}$, i.e. elliptic Nekrasov factors. In section \ref{s:ellipticAFS}, we will show that the elliptic vertex introduced there reproduces the above instanton partition function (elliptic genus) by reproducing the correct expressions of $Z_{bfd}^{ellip}$ and $Z_{vec}^{ellip}$. 

\section{Elliptic Ding-Iohara-Miki Algebra \cite{Saito}}

In the following sections, we construct an elliptic topological vertex, which is formulated on the toric diagram (brane web before compactification) while reproduces the elliptic partition function. We start from the definition and basic properties of the elliptic DIM algebra. 

{\bf Definition 1} The elliptic DIM algebra is specified by the generators $\psi^\pm(z)$, $x^{\pm}(z)$, and the following relations ($q_1q_2q_3=1$),
\begin{eqnarray}
[\psi^\pm (z), \psi^\pm (w)] & = & 0,
\\
\psi^+ (z) \, \psi^- (w) & = & 
\frac{
g_{ellip} \lt( \gammah      z/w \rt) 
}{
g_{ellip} \lt( \gammah^{-1} z/w \rt) 
} 
\psi^- (w) \, \psi^+ (z)
\label{pp}
\\
\psi^\pm (z) \, x^+ (w) & = & g_{ellip} \lt( \gammah^{\pm \frac12} z/w \rt)    x^+ (w) \, \psi^\pm (z)
\label{px+}
\\
\psi^\pm (z) \, x^- (w) & = & g_{ellip} \lt( \gammah^{\mp \frac12} z/w \rt)^{-1} x^-(w) \, \psi^\pm (z)
\label{px-}
\\
x^\pm (z) \, x^\pm (w)  & = & g_{ellip} \lt( z/w \rt)^{\pm 1} x^\pm (w) \, x^\pm (z)
\label{xx1}
\\ 
\left[x^+(z), x^-(w)\right]  & = & 
\frac{\theta_p (q_1) \theta_p (q_2)}{ \lt( p; p \rt)^2_\infty \theta_p (q_1 q_2)}
\lt( \delta \lt( \gammah      w/z \rt)  \, \psi^+ \lt( \gammah^{  \frac12} w \rt) - 
    \delta \lt( \gammah^{-1} w/z \rt)  \, \psi^- \lt( \gammah^{- \frac12} w \rt)  
\rt), 
\label{xx2}
\end{eqnarray}
with, 
\ba
g_{ellip}(z)=\prod_i\frac{\theta_p(q_iz)}{\theta_p(q_i^{-1}z)}.
\ea

{\bf Remark:} There is an apparent automorphism of the algebra, i.e. interchanging among the parameters $q_1\leftrightarrow q_2\leftrightarrow q_3$. We set, when no further instruction is added, $q_1=q$, $q_2=t^{-1}$ and $q_3=q_1^{-1}q_2^{-1}=t/q$ in this article. 

This algebra admits a Hopf algebra structure, with a standard Drinfeld coproduct, as can be already seen in the discussion of Ding and Iohara's original paper \cite{DI}. We have the following lemma for the concrete form of the coproduct and it is shown by explicit computation in Appendix \ref{a-coproduct}. 

{\bf Lemma 1} The elliptic DIM algebra is equipped with the following coproduct structure. 
\begin{align}
\begin{split}\label{AFS_coproduct}
&\Delta(x^+(z))=x^+(z)\otimes 1+\psi^-(\hat{\gamma}_{(1)}^{1/2}z)\otimes x^+(\hat{\gamma}_{(1)}z),\\
&\Delta(x^-(z))=x^-(\hat{\gamma}_{(2)} z)\otimes \psi^+(\hat{\gamma}_{(2)}^{1/2}z)+1\otimes x^-(z),\\
&\Delta(\psi^+(z))=\psi^+(\hat{\gamma}_{(2)}^{1/2}z)\otimes\psi^+(\hat{\gamma}_{(1)}^{-1/2}z),\\
&\Delta(\psi^-(z))=\psi^-(\hat{\gamma}_{(2)}^{-1/2}z)\otimes\psi^-(\hat{\gamma}_{(1)}^{1/2}z),
\end{split}
\end{align}
where $\hat{\gamma}_{(1)}=\hat{\gamma}\otimes 1$ and $\hat{\gamma}_{(2)}=1\otimes \hat{\gamma}$. 

\begin{flushright}
$\Box$
\end{flushright}

{\bf Remark:} We note that from $\theta_p(z^{-1})=-z^{-1}\theta_p(z)$, an important property, 
\ba
g_{ellip}(z^{-1})=g_{ellip}(z)^{-1},
\ea
holds for the algebra. 

{\bf Remark:} Unlike in the usual DIM, the mode expansion of $x^\pm(z)$ and $\psi^\pm(z)$ are all given by \cite{Saito}
\ba
x^\pm(z)=:\sum_{n\in\mathbb{Z}}x^\pm_n z^{-n},\quad \psi^\pm(z)=:\sum_{n\in\mathbb{Z}}\psi^\pm_nz^{-n},
\ea
the direct consequence of which is that $\psi^\pm_0$ are no longer centers of the algebra. 

\section{Representations of Elliptic DIM}

In this section, we briefly discuss several representations of the elliptic DIM, in analogy to the vertical and horizontal representations of the usual DIM algebra. These representations will be used in the construction of the elliptic vertex in the next section. An important feature of these representations is that in the $p\rightarrow 0$ limit, they reduce to well-defined and yet non-trivial representations of the usual DIM algebra, that is to say, if we further expand the generators of the elliptic DIM with respect to $p$, as 
\ba
x^\pm(z)=:\sum_{n\in\mathbb{Z},m\in\mathbb{Z}_{\geq0}}x^\pm_{m,n} z^{-n}p^m,\quad \psi^\pm(z)=:\sum_{n\in\mathbb{Z},m\in\mathbb{Z}_{\geq0}}\psi^\pm_{m,n}z^{-n}p^m,\label{double-expansion}
\ea
$\psi^\pm_{0,0}$ are mapped to constants in these representations (since they are central elements in  DIM), and give rise to an additional label to $\gammah$ to this class of representations.

\subsection{Vertical Representation}\label{s:vert-rep}

Recall that the $(0,N)$ vertical representation of the normal DIM reads 
\ba
&&x^+(z)\ket{\vec{v},\vec{\lambda}}\rangle \propto z^{-N+1}\sum_{x\in A(\vec{\lambda})}\delta(z/\chi_x)\Res_{z\rightarrow \chi_x}\frac{1}{z\cY_{\vec{\lambda}}}\ket{\vec{v},\vec{\lambda}+x}\rangle,\nn\\
&&x^-(z)\ket{\vec{v},\vec{\lambda}}\rangle\propto z^{N-1}\sum_{x\in R(\vec{\lambda})}\delta(z/\chi_x)\Res_{z\rightarrow\chi_x}z^{-1}\cY_{\vec{\lambda}}(zq_3^{-1})\ket{\vec{v},\vec{\lambda}-x}\rangle,\nn\\
&&\psi^\pm(z)\ket{\vec{v},\vec{\lambda}}\rangle\propto\Psi_{\vec{\lambda}}(z)\ket{\vec{v},\vec{\lambda}}\rangle\nn,
\ea
which preserves the norm $\langle\langle \vec{v},\vec{\lambda}\ket{\vec{v},\vec{\mu}}\rangle=\delta_{\vec{\lambda},\vec{\mu}}a^{-1}_{\vec{\lambda}}$ with a natural dual representation.  A similar representation labeled by one Young diagram of the elliptic DIM which preserves the elliptic instanton partition function can be constructed. The existence of this representation is the key to build the elliptic vertex. 

{\bf Proposition 1} The following action of generators on the basis labeled by the highest-weight parameter $v$ and a Young diagram $\lambda$ gives a representation of the elliptic DIM algebra at $\gammah=1$. 
\ba
&&x^+(z)\ket{v,\lambda}\rangle=\frac{b}{(p;p)^2_\infty} \sum_{x\in A(\lambda)}\delta(z/\chi_x)\Res_{z\rightarrow \chi_x}{}^\theta\frac{1}{z\cY^{ellip}_\lambda(z)}\ket{v,\lambda+x}\rangle,\\
&&x^-(z)\ket{v,\lambda}\rangle=\frac{b^{-1}\gamma^{-1}}{(p;p)^2_\infty} \sum_{x\in R(\lambda)}\delta(z/\chi_x)\Res_{z\rightarrow \chi_x}{}^\theta z^{-1}\cY^{ellip}_\lambda(zq_3^{-1})\ket{v,\lambda-x}\rangle,\\
&&\psi^\pm(z)\ket{v,\lambda}\rangle=\gamma^{-1}\Psi^{ellip}_\lambda(z)\ket{v,\lambda}\rangle,
\ea
where the elliptic $\cY$-function is given by 
\ba
\cY^{ellip}_\lambda(z)=\frac{\prod_{x\in A(\lambda)}\theta_p(\chi_x/z)}{\prod_{x\in R(\lambda)}\theta_p(\chi_x/(zq_3))},\label{def-Y}
\ea
\ba
\Psi^{ellip}_\lambda(z)=\cY^{ellip}_\lambda(zq^{-1}_3)/\cY^{ellip}_\lambda(z),
\ea 
and we introduced a notation 
\ba
\Res_{z\rightarrow x}{}^\theta f(z)=\frac{1}{(p;p)_\infty^2}\Res_{z\rightarrow x}f(z),\label{ellip-Res}
\ea
as an elliptic residue. $A(\lambda)$ and $R(\lambda)$ are respectively the set of boxes that can be added to or removed from the Young diagram $\lambda$ to create a Young diagram with size $|\lambda|\pm 1$. $b$ is a free parameter here. We call this representation the vertical representation of the elliptic DIM algebra. 
\begin{flushright}
$\Box$
\end{flushright}

The proof is due to the explicit  computation and we present it in Appendix \ref{vert-rep}. 

There is one important point to stress here. It may appear in the above expression of the vertical representation that the representations of $\psi^+(z)$ and $\psi^-(z)$ are the same, but they are in fact different expansion of the same function. In the calculation of Appendix \ref{vert-rep}, to establish the representation with the generalized Ramanujan's identity (\ref{Ramanujan-identity}), we replace $\theta$-functions in $\psi^+$, $\theta_p(z)$, with the identity $\theta_p(z)=-z\theta_p(z^{-1})$ to obtain the expression of $\psi^-$. Note that 
\ba
\theta_p(z)=(1-z)\prod_{i=1}^\infty(1-p^iz)(1-p^iz^{-1}),\label{theta-split}
\ea
is symmetric about $z\leftrightarrow z^{-1}$ in the infinite product, and thus replacing $\theta_p(z)$ with $\theta_p(z)=-z\theta_p(z^{-1})$ is essentially replacing $(1-z)$ with $-\frac{(1-z^{-1})}{z^{-1}}$. When a $\theta_p(z)$ is in the denominator, this operation gives rise to a difference proportional to the $\delta$-function. 

The simplest way to work out the double expansion of $\psi^\pm(z)$, (\ref{double-expansion}), is to first split the $\theta$-function as (\ref{theta-split}) into an infinite product, which is symmetric about $z\leftrightarrow z^{-1}$, and the part that survives in the $p\rightarrow 0$ limit (we call it the DIM part), and we expand the infinite product with repsect to $p$ in the same way as the well-known expansion of $\theta_p(z)$ presented in (\ref{expand-theta}) and the DIM part about $z$ (resp. $z^{-1}$) for $\psi^+$ ($\psi^-$ resp.). This difference on expansion can be viewed as inherited from the representation of the usual DIM algebra obtained in the $p\rightarrow 0$ limit, as the only difference between $\psi^\pm$ comes from the factors in the DIM part. 

We have 
\ba
\psi^\pm_{0,0}=\gamma^{\mp1},
\ea
in this representation.

{\bf Remark:} We fixed the overall scaling of $x^\pm$ in the above expressions, but left the relative scaling $b$ to be free for now. It turns out that there is a nice and natural choice of $b$ for a particular expression of the elliptic vertex. 

To construct a dual vertical representation, we need to fix $a_{\lambda}$, the inner product of two dual states, in the elliptic case. Recall that in the usual DIM, we can choose 
\ba
a_{\vec{\lambda}}=Z_{vect}(\vec{v},\vec{\lambda})\prod_{l=1}^N(\gamma v_l)^{-|\vec{\lambda}|}\prod_{x\in\vec{\lambda}}\chi_x,
\ea
then the most natural candidate in the elliptic case is obtained by simply replacing $Z_{vect}$ with its elliptic version, (\ref{vec-ellip}), using 
\ba
Z^{ellip}_{bfd}(v,\lambda;v',\nu|\mu)=\prod_{(i,j)\in\lambda}\theta_p\lt((v/v')\mu^{-1}q^{-\nu^t_j+i}t^{-\lambda_i+j-1}\rt)\prod_{(i,j)\in\nu}\theta_p\lt((v/v')\mu^{-1}q^{\lambda^t_j-i+1}t^{\nu_i-j}\rt),
\ea
i.e. 
\ba
a^{ellip}_{\lambda}=Z^{ellip}_{vect}(v,\lambda)(\gamma v)^{-|\lambda|}\prod_{x\in\lambda}\chi_x.
\ea
Using the recursive relations of $Z^{ellip}_{vect}(v,\lambda)$ presented in Appendix \ref{Nekra}, we have 
\ba
&&\frac{a^{ellip}_{\lambda+x}}{a^{ellip}_\lambda}=\frac{\theta_p(q_3^{-1})}{\theta_p(q_1)\theta_p(q_2)}\gamma\chi_x^{-1}\Res_{z\rightarrow \chi_x}{}^\theta\frac{1}{\cY_\lambda(zq_3^{-1})\cY_\lambda(z)},\label{a+}\\
&&\frac{a^{ellip}_{\lambda-x}}{a^{ellip}_\lambda}=-\frac{\theta_p(q_3^{-1})}{\theta_p(q_1)\theta_p(q_2)}\gamma^{-1}\chi_x^{-1}\Res_{z\rightarrow \chi_x}{}^\theta\cY_\lambda(zq_3^{-1})\cY_\lambda(z),\label{a-}
\ea
where we used the identity $\theta_p(z^{-1})=-z^{-1}\theta_p(z)$. 

With (\ref{a+}) and (\ref{a-}), it is straightforward to derive the following corollary. 

{\bf Corollary 1} The following dual vertical representation, 
\ba
&&\langle\bra{v,\lambda}\psi^\pm(z)=\gamma^{-1}\Psi^{ellip}_\lambda(z)\langle\bra{v,\lambda},\\
&&\langle \bra{v,\lambda}x^+(z)=-\frac{\gamma^{-1}b}{(p;p)^2_\infty}\sum_{x\in R(\lambda)}\langle\bra{v,\lambda-x}\delta(z/\chi_x)\Res_{z\rightarrow \chi_x}{}^\theta z^{-1}\cY^{ellip}_\lambda(zq_3^{-1}),\\
&&\langle\bra{v,\lambda}x^-(z)=-\frac{b^{-1}}{(p;p)^2_\infty}\sum_{x\in A(\lambda)}\langle \bra{v,\lambda+x}\delta(z/\chi_x)\Res_{z\rightarrow \chi_x}{}^\theta \frac{1}{z\cY^{ellip}_\lambda(z)},
\ea
admits a norm structure $\langle\langle v,\lambda\ket{v,\mu}\rangle=\delta_{\lambda,\mu}\lt(a^{ellip}_\lambda\rt)^{-1}$. 

{\bf Remark:} We will denote the map from elements of the elliptic DIM algebra to operators in the vertical representation space by $\rho^{(0,1)}_v$. Its action on the ket (resp. bra) basis is given by the vertical (resp. dual vertical) representation described above. The corresponding representation space is denoted as ${\cal F}^{(0,1)}_v$. The label $(\ell_1,\ell_2)=(0,1)$ is inherited from the representation of DIM, and in the current case $\rho^{(\ell_1,\ell_2)}_v(\gammah,\psi^+_{0,0}/\psi^-_{0,0})=(q_3^{\ell_1/2},q_3^{-\ell_2})$.

\subsection{Horizontal representations \cite{Saito}}\label{s:hori-rep}

We completely follow the convention used in \cite{Saito} in this subsection. 

{\bf Definition 2 (elliptic boson)} The elliptic boson is constituted of two copies of Heisenberg algebras with the normalization, 
\ba
\lt[a_m,a_n\rt]=m(1-p^{|m|})\frac{1-q^{|m|}}{1-t^{|m|}}\delta_{m+n,0},\quad \lt[b_m,b_n\rt]=m\frac{1-p^{|m|}}{(qt^{-1}p)^{|m|}}\frac{1-q^{|m|}}{1-t^{|m|}}\delta_{m+n,0},
\ea
and $a_n$ and $b_n$ commute with each other. 

{\bf Definition 3} We define 
\ba
&&\eta(z)=:\exp\lt(-\sum_{n\neq 0}\frac{1-t^{-n}}{1-p^{|n|}}p^{|n|}b_n\frac{z^n}{n}\rt)\exp\lt(-\sum_{n\neq 0}\frac{1-t^n}{1-p^{|n|}}a_n\frac{z^{-n}}{n}\rt):,\label{def-eta}\\
&&\xi(z)=:\exp\lt(\sum_{n\neq 0}\frac{1-t^{-n}}{1-p^{|n|}}\gamma^{-|n|}p^{|n|}b_n\frac{z^n}{n}\rt)\exp\lt(\sum_{n\neq 0}\frac{1-t^n}{1-p^{|n|}}\gamma^{|n|}a_n\frac{z^{-n}}{n}\rt):,\label{def-xi}
\ea
\ba
&&\varphi(z)=:\exp\lt(-\sum_{n>0}\frac{(1-t^n)(qt^{-1}p)^n}{1-p^n}(1-\gamma^{2n})\gamma^{n/2}b_{-n}\frac{z^{-n}}{n}\rt)\exp\lt(\sum_{n>0}\frac{1-t^{-n}}{1-p^n}(1-\gamma^{2n})\gamma^{-n/2}a_{-n}\frac{z^n}{n}\rt):,\nn\\
&&\label{def-phi}\\
&&\bar{\varphi}(z)=:\exp\lt(\sum_{n>0}\frac{(1-t^{-n})(qt^{-1}p)^n}{1-p^n}(1-\gamma^{2n})\gamma^{n/2}b_n\frac{z^{n}}{n}\rt)\exp\lt(-\sum_{n>0}\frac{1-t^n}{1-p^n}(1-\gamma^{2n})\gamma^{-n/2}a_n\frac{z^{-n}}{n}\rt):.\nn\\
&&\label{def-bphi}
\ea

As shown in \cite{Saito} we have the following representation of the elliptic DIM labeled by $n$. 

{\bf Proposition 2} The following map gives a representation of the elliptic DIM algebra at $\gammah=\gamma$, 
\ba
x^+(z)\mapsto u\gamma^{n}z^{-n}\eta(z),\quad x^-(z)\mapsto u^{-1}\gamma^{-n}z^{n}\xi(z),\quad \psi^+(z)\mapsto \gamma^{-n}\bar\varphi(z),\quad \psi^-(z)\mapsto \gamma^n\varphi(z).\label{hor-rep-n}
\ea
$u$ is the highest-weight parameter of the representation. We denote this map with the notation $\rho^{(1,n)}_u$, and ${\cal F}^{(1,n)}_u$ as the representation space of it\footnote{${\cal F}^{(1,n)}_u$ for any $n$ and $u$ is essentially a tensor product of two Fock spaces.}. 
\begin{flushright}
$\Box$
\end{flushright}

We check this claim in Appendix \ref{hor-rep}. 

{\bf Remark:} In the definition of vertex operators (\ref{def-eta})-(\ref{def-bphi}), we observe that whenever $b_n$ appear, it is always accompanied by a factor $p^{|n|}$. We can absorb this factor into the normalization of $b_n$, and then the Heisenberg algebra of $b_n$ becomes completely commutative in the limit $p\rightarrow 0$. In this limit, $b_n$ no longer contributes any factor to the correlators, and can be completely dropped out. Therefore, the vertex operators (\ref{def-eta})-(\ref{def-bphi}) reduce to those used in the $(1,n)$ representation of DIM algebra \cite{FHHSY, AFS}, when we take $p\rightarrow 0$. 

{\bf Remark:} Note that $\psi^\pm_{0,0}=\gamma^{\mp n}$, and indeed $(\ell_1,\ell_2)=(1,n)$ for the representation (\ref{hor-rep-n}). 

\section{Elliptic Awata-Feigin-Shiraishi vertex}\label{s:ellipticAFS}

As explained in \cite{BFHMZ}, in the AFS approach, each D5-brane (or $(1,0)$-brane) in the brane web is mapped to a vertical representation in the algebra and each $(n,1)$-brane to a $(1,n)$ horizontal representation. The highest weight parameters $u$'s and $v$'s respectively have the physical meaning of the position parameters of NS5 and D5-branes. At the intersections of three branes, we assign AFS vertices to the brane web, and the VEV of the product of all these vertices gives the instanton partition function of the brane web. In the elliptic case, we will use the elliptic vertex introduced in the this section instead, and will adopt the same rule of assignment to the brane web (without compactification). 

{\bf Definition 4 (elliptic vertex)} We introduce the following elliptic vertices 
\ba
\Phi^{(n)}:{\cal F}^{(1,n)}_u\otimes {\cal F}^{(0,1)}_v\rightarrow {\cal F}^{(1,n+1)}_{-uv},\\
\Phi^{\ast(n)}:{\cal F}^{(1,n+1)}_{-uv}\rightarrow {\cal F}^{(1,n)}_u\otimes {\cal F}^{(0,1)}_v,
\ea
and define them in terms of their vector components, 
\ba
\Phi^{(n)}_\lambda[u,v]:=\Phi^{(n)}[u,v]\ket{v,\lambda}\rangle=t_n(\lambda,u,v):\Phi_\emptyset(v)\prod_{x\in\lambda}\eta(\chi_x):,\label{def-Phi}\\
\Phi^{\ast(n)}_\lambda[u,v]:=\langle\bra{v,\lambda}\Phi^{\ast(n)}[u,v]=t^\ast_n(\lambda,u,v):\Phi^\ast_\emptyset(v)\prod_{x\in\lambda}\xi(\chi_x):\label{def-Phis},
\ea
where $t_n$ and $t^\ast_n$ are again given by 
\ba
t_n(\lambda,u,v)=(-uv)^{|\lambda|}\prod_{x\in\lambda}(\gamma/\chi_x)^{n+1},\\
t^\ast_n(\lambda,u,v)=(\gamma u)^{-|\lambda|}\prod_{x\in\lambda}(\chi_x/\gamma)^n,
\ea
and 
\ba
\Phi_\emptyset(v)=:\exp\lt(-\sum_{n\neq0}\frac{1}{n}\frac{q^{-n}v^{n}}{(1-q^{-n})(1-p^{|n|})}p^{|n|}b_n\rt)\exp\lt(\sum_{n\neq 0}\frac{1}{n}\frac{v^{-n}}{(1-q^{-n})(1-p^{|n|})}a_{n}\rt):,\\
\Phi_\emptyset^\ast(v)=:\exp\lt(\sum_{n\neq0}\frac{1}{n}\frac{\gamma^{-|n|}q^{-n}v^{n}}{(1-q^{-n})(1-p^{|n|})}p^{|n|}b_n\rt)\exp\lt(-\sum_{n\neq 0}\frac{1}{n}\frac{\gamma^{|n|} v^{-n}}{(1-q^{-n})(1-p^{|n|})}a_{n}\rt):.
\ea
\begin{flushright}
$\Box$
\end{flushright}

{\bf Proposition 3} The contraction between elliptic vertices can be computed to the following results. 
\ba
&&\contraction{}{\Phi_\mu[u_2,v_2]}{}{\Phi_\lambda[u_1,v_1]}\Phi_\mu[u_2,v_2]\Phi_\lambda[u_1,v_1]=\tilde{\cal G}^{-1}(v_1,v_2)Z_{bfd}^{ellip}(v_1,v_2;\lambda,\mu|1)^{-1}:\Phi_\mu[u_2,v_2]\Phi_\lambda[u_1,v_1]:,\label{contra-1}\\
&&\contraction{}{\Phi^\ast_\mu[u_2,v_2]}{}{\Phi^\ast_\lambda[u_1,v_1]}\Phi^\ast_\mu[u_2,v_2]\Phi^\ast_\lambda[u_1,v_1]=\tilde{\cal G}^{-1}(v_1,v_2q_3^{-1})Z_{bfd}^{ellip}(v_1,v_2q_3^{-1};\lambda,\mu|1)^{-1}:\Phi^\ast_\mu[u_2,v_2]\Phi^\ast_\lambda[u_1,v_1]:,\label{contra-2}
\ea
\ba
\contraction{}{\Phi_\mu[u_2,v_2]}{}{\Phi^\ast_\lambda[u_1,v_1]}\Phi_\mu[u_2,v_2]\Phi^\ast_\lambda[u_1,v_1]=\tilde{\cal G}(v_1,v_2\gamma^{-1})Z^{ellip}_{bfd}(v_1,v_2\gamma^{-1};\lambda,\mu|1):\Phi_\mu[u_2,v_2]\Phi^\ast_\lambda[u_1,v_1]:,\label{contra-3}\\
\contraction{}{\Phi^\ast_\mu[u_2,v_2]}{}{\Phi_\lambda[u_1,v_1]}\Phi^\ast_\mu[u_2,v_2]\Phi_\lambda[u_1,v_1]=\tilde{\cal G}(v_1,v_2\gamma^{-1})Z^{ellip}_{bfd}(v_1,v_2\gamma^{-1};\lambda,\mu|1):\Phi^\ast_\mu[u_2,v_2]\Phi_\lambda[u_1,v_1]:\label{contra-4}.
\ea
where 
\ba
\tilde{{\cal G}}(v_1,v_2)=\exp\lt(\sum_{n>0}\frac{1}{n}\frac{(v_1/v_2)^n}{(1-t^n)(1-q^{-n})(1-p^n)}\rt)\exp\lt(\sum_{n>0}\frac{1}{n}\frac{p^n}{1-p^n}\frac{q_3^n(v_2/v_1)^n}{(1-q^{-n})(1-t^n)}\rt),
\ea
and we used the identity (\ref{AFS-Nekra}), 
\ba
Z^{ellip}_{bfd}(v_1,v_2;\lambda,\mu)=\prod_{x\in\lambda}\theta_p\lt(\frac{\chi_x}{v_2q_3}\rt)\prod_{y\in\mu}\theta_p\lt(\frac{v_1}{\chi_y}\rt)\prod_{x\in\lambda,y\in\mu}\frac{\theta_p(q\chi_x\chi_y^{-1})\theta_p(t^{-1}\chi_x\chi_y^{-1})}{\theta_p(\chi_x\chi_y^{-1})\theta_p(\chi_x\chi_y^{-1}q_3^{-1})}.
\ea
\begin{flushright}
$\Box$
\end{flushright}

{\bf Remark:} We note that up to a factor independent of $v_1$ and $v_2$, $\tilde{\cal G}(v_1,v_2)$ matches the overall factor ${\cal G}^{ellip}(v_1,v_2;q,t,p)$ derived in (\ref{G-ellip}).

{\bf Corollary 2} With the above properties (\ref{contra-1}), (\ref{contra-2}), (\ref{contra-3}), (\ref{contra-4}) of the elliptic vertices, we reproduce the elliptic genus of M-strings computed for example in \cite{ellip-Virasoro}. 

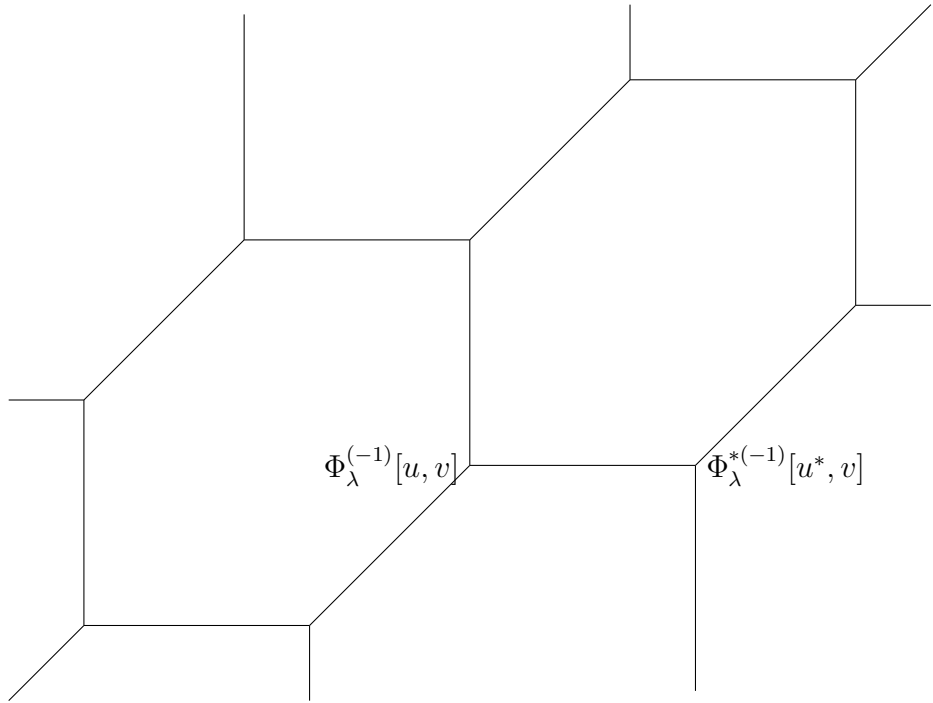
\begin{figure}
\begin{center}
\begin{tikzpicture}
      \draw (0,0)--(3,0);
      \draw (0,0)--(0,3);
      \draw (0,0)--(-2.13,-2.13);
      \draw (3,0)--(5.13,2.13);
      \draw (3,0)--(3,-3);
      \draw (5.13,2.13)--(5.13,3.13);
      \draw (5.13,2.13)--(8.13,2.13);
      \draw (3,-3)--(0.87,-5.13);
      \draw (0.87,-5.13)--(-2.13,-5.13);
      \draw (-2.13,-2.13)--(-2.13,-5.13);
      \draw (3,-3)--(6,-3);
      \draw (6,-3)--(8.13,-0.87);
      \draw (8.13,-0.87)--(8.13,2.13);
      \draw (8.13,2.13)--(9.13,3.13);
      \draw (8.13,-0.87)--(9.13,-0.87);
      \draw (6,-3)--(6,-6);
      \draw (0.87,-5.13)--(0.87,-6.13);
      \draw (-2.13,-5.13)--(-3.13,-6.13);
      \draw (-2.13,-2.13)--(-3.13,-2.13);
      \node at (3,-3) [left] {$\Phi^{(-1)}_\lambda[u,v]$};
      \node at (6,-3) [right] {$\Phi^{\ast(-1)}_\lambda[u^\ast,v]$};
      \end{tikzpicture}
      \caption{A typical representation web (toric diagram) for M-strings.}
      \label{ellip-web}
\end{center}
\end{figure}

{\bf Proof:} What needs to be reproduced is the elliptic Nekrasov factor (\ref{ellip-bifund}) from the contraction of vertex operators in the horizontal representation and (\ref{vec-ellip}) from the vertical representation. We recall how the AFS vertices are glued together: they are simply multiplied together according to the brane web, and we take the vacuum expectation value of the multiplied object to obtain the partition function. In the vertical representation, in order to use the concrete expressions (\ref{def-Phi}) and (\ref{def-Phis}), we insert the identity operator ${\bf id}=\sum_{\lambda}\ket{v,\lambda}\rangle a^{ellip}_\lambda \langle \bra{v,\lambda}$ into ${\cal F}^{(0,1)}_v$, and (\ref{vec-ellip}) is reproduced from (see Figure \ref{ellip-web})
\ba
t_{-1}(\lambda,u,v)t^\ast_{-1}(\lambda,u^\ast,v)a^{ellip}_\lambda=\lt(-\frac{u}{u^\ast\gamma}\rt)^{|\lambda|}Z^{ellip}_{vect}(v,\lambda),
\ea
where $\mathfrak{q}:=-\frac{u}{u^\ast\gamma}$ is the gauge coupling associated to ${\cal F}^{(0,1)}_v$. The remaining part is trivial from the contraction (\ref{contra-1}), (\ref{contra-2}), (\ref{contra-3}) and (\ref{contra-4}). Refer to, for example, \cite{BFHMZ} or \cite{D-type} for how to construct an instanton partition with a given shape of quiver. 
\begin{flushright}
$\Box$
\end{flushright}

The main claim of this paper is that the elliptic vertex defined above not only reproduces the elliptic genus of M-strings, but is also an important building element of the underlying elliptic DIM algebra, i.e. the intertwiner of the algebra. 

{\bf Theorem (AFS property)} The AFS property of the intertwiner, 
\ba
\Phi^{(n)}[u,v](\rho^{(0,1)}_{v}\otimes \rho^{(1,n)}_u)\Delta(g(z))=\rho^{(1,n+1)}_{u'}(g(z))\Phi^{(n)}[u,v],\label{AFS-prop-1}\\
(\rho^{(1,n)}_{u}\otimes \rho^{(0,1)}_{v})\Delta(g(z))\Phi^{\ast(n)}[u,v]=\Phi^{\ast(n)}[u,v]\rho^{(1,n+1)}_{-uv}(g(z)).\label{AFS-prop-2}
\ea
with $g=x^\pm,\ \psi^\pm$, holds for elliptic vertices. 
\begin{flushright}
$\Box$
\end{flushright}

The proof is given in Appendix \ref{AFS-prop} by explicit computation.

\section{Conclusion $\&$ Discussion}

In this article, we constructed a vertical representation whose basis is labeled by one Young diagram for the elliptic DIM algebra at $\gammah=1$, and based on it, we did a parallel work to \cite{AFS} to build an elliptic vertex, whose VEV associated to certain class of representation webs (see Figure \ref{ellip-web}) reproduces the partition function of corresponding 6d $\cN=(1,0)$ SCFTs (or equivalently, elliptic genus of M-strings) up to some factor independent of Coulomb branch (or K\"ahler) parameters. There are, however, still a lot of open questions for future works, and we list some of them here with brief and limited discussions attached. 

\paragraph{Iqbal-Kozcaz-Vafa analogue of the elliptic vertex} We note that essentially what is done in this article is to replace the trace of a free boson with the VEV of two copies of free bosons (elliptic boson) for the AFS refined topological vertex, and this is known as the Clavelli-Shapiro \cite{Clavelli-Shapiro} trick (see also \cite{FHSY-CS} for a brief review) in the literature. In the IKV version of the refined topological vertex, the bifundamental factors obtained in the unpreferred direction (which corresponds to the horizontal representation direction in the AFS approach) can also be viewed as VEVs of vertex operators by using the vertex operator expression of the skew Schur function 
\ba
s_{\mu/\nu}(\vec{x})=\bra{\nu}V_+(\vec{x})\ket{\mu}=\bra{\mu}V_-(\vec{x})\ket{\nu},\label{skew-schur}
\ea
where $\ket{\mu}$, $\ket{\nu}$ are the fermion basis associated to the Frobenius coordinate of Young diagrams $\mu$, $\nu$ (see (\ref{Frobenius}) and \cite{IKV}) and 
\ba
J_n:=\sum_{j\in \mathbb{Z}+1/2}\psi_{-j}\psi^\ast_{j+n},\quad V_\pm (\vec{x})=\exp\lt(\sum_{n=1}^\infty \frac{1}{n}\sum_i x_i^n J_{\pm n}\rt).
\ea
We can also apply the Clavelli-Shapiro trick to these vertex operators and work out the IKV version of the elliptic vertex, or we can follow the discussion in \cite{AFS} to work the IKV version as the matrix element of the elliptic vertex introduced in this article . The details are reported in \cite{Foda-Zhu}. 

\paragraph{Representations of elliptic DIM}

As we have already remarked in section \ref{s:vert-rep} and \ref{s:hori-rep}, all the representations of the elliptic DIM algebra we considered in this article have the special property $\psi^\pm_{0,0}={\rm constant}$. This property is inherited through the coproduct, due to the fact that the expansion only contains non-negative powers of $p$. In this paper, it follows from the fact that in the $p\rightarrow 0$ limit the representation we consider always reduces to a representation of the DIM algebra. It seems that representations with $\psi^\pm_{0,0}$ central might only be a subclass of all representations of the elliptic DIM algebra. We do not know any concrete example of more general representations, which does not satisfy this property, and if there exists such an example, it will be curious what kind of role it can play in the classification of 6d gauge theories. 

\paragraph{Generalized elliptic vertex} It is possible to write down a vertical representation with $\psi^\pm_{0,0}=\gamma^{\mp n}$ labeled by $n$ Young diagrams for the elliptic DIM algebra at $\gammah=1$. This can be done simply by replacing the $\cY$-function with 
\ba
\cY^{ellip}_{\vec{\lambda}}(z)=\frac{\prod_{x\in A(\vec{\lambda})}\theta_p(\chi_x/z)}{\prod_{x\in R(\vec{\lambda})}\theta_p(\chi_x/(zq_3))},
\ea
where for $\vec{\lambda}=(\lambda^{(1)},\lambda^{(2)},\dots,\lambda^{(n)})$, $A(\vec{\lambda})=A(\lambda^{(1)})\cup A(\lambda^{(2)})\cup \dots \cup A(\lambda^{(n)})$ and $R(\vec{\lambda})=R(\lambda^{(1)})\cup R(\lambda^{(2)})\cup \dots \cup R(\lambda^{(n)})$, and properly adjusting the prefactors to symmetrize the expression for $x^\pm$ and $\psi^\pm$. With this representation, we can construct the generalized elliptic vertex as done in \cite{BFHMZ}, and this generalized vertex simplifies the computation for higher rank gauge groups a lot, allowing us to make general statements.

\paragraph{qq-character of Kimura-Pestun's picture}

Since the AFS property, (\ref{AFS-prop-1}), (\ref{AFS-prop-2}), holds for the elliptic vertex introduced in this article, a parallel analysis to that done in \cite{BFHMZ} can be repeated to construct quantities in the algebra commutating with the ${\cal T}$-operator, which is product of $\Phi$ and $\Phi^\ast$ and whose VEV gives the instanton partition function of the corresponding brane web. These quantities are argued to correspond to Kimura-Pestun's generators \cite{Kimura-Pestun} of the quiver $\cW$-algebra, $\cW_{q,t}(\Gamma)$. The generalization of \cite{Kimura-Pestun} to 6d is presented in \cite{ellip-Kimura-Pestun}. We note that these generators take almost the same form as their siblings in 5d, and resemble the structure of the qq-character generators built from ``surface" operators in \cite{KMS}. It will be very interesting to find the correspondings of the ``surface" operators in the elliptic DIM algebra. 

\paragraph{qq-character in vertical representation and constrains on matter contents} Another way to derive the qq-characters from the DIM algebra was proposed in \cite{5dBMZ}, in which we convert the Ward identity of elements of the DIM algebra acting on the brane web to the contour integral, e.g. for a pure SU($n$) gauge theory, 
\ba
\oint\frac{{\rm d}x}{2\pi i}\frac{1}{(x-z)}\lt\langle x^n\cY(xq_3^{-1})+\frac{\mathfrak{q}\tilde{\nu}^{-1}}{\cY(x)}\rt\rangle,
\ea
where $\tilde{\nu}=\prod_{i=1}^n\lt(-\frac{1}{v_iq_3}\rt)$. The Ward identity states that the sum of all residues over $\{\chi_x|x\in A(\lambda)\cup R(\lambda)\}$ when evaluated on $\ket{\vec{v},\vec{\lambda}}\rangle$ in the expectation value 
\ba
\langle \dots\rangle=\frac{\sum_\lambda t_{-1}(\lambda,u,v)t^\ast_{-1}(\lambda,u^\ast,v)a_\lambda \langle\bra{\vec{v},\vec{\lambda}}\dots\ket{\vec{v},\vec{\lambda}}\rangle}{\sum_\lambda t_{-1}(\lambda,u,v)t^\ast_{-1}(\lambda,u^\ast,v)a_\lambda \langle\bra{\vec{v},\vec{\lambda}}{\bf id}\ket{\vec{v},\vec{\lambda}}\rangle},
\ea
is zero. 
$\cY(x)$ is an operator not specified in this article, and we only need its property that $\langle\bra{\vec{v},\vec{\lambda}} \cY(x)\ket{\vec{v},\vec{\lambda}}\rangle =\cY_{\vec{\lambda}}(x)\langle\langle \vec{v},\vec{\lambda}\ket{\vec{v},\vec{\lambda}}\rangle$. At the end, we find that the integral around $x\sim z$ can be expressed as the sum of residues around $x\sim 0$ and $x\sim\infty$. It turns out that 
\ba
\lt\langle z^n\cY(zq_3^{-1})+\frac{\mathfrak{q}\tilde{\nu}^{-1}}{\cY(z)}\rt\rangle=T_n(z),
\ea
with $T_n(z)$ some polynomial of $z$ of degree $n$. We can also add additional matter flavors and non-trivial Chern-Simons level to the theory, and as discussed in \cite{5dBMZ}, by requiring that $T_n(z)$ remains to be a polynomial of degree $n$, we can recover the constraint on number of flavors and Chern-Simons level for 5d SU($n$) originally derived in \cite{Intriligator-Morrison-Seiberg}. 

We can repeat the same computation with the elliptic vertices, and interestingly, we see that if either the number of fundamental hypermultiplets or the number of anti-fundamental hypermultiplets is not equal to $n$ in SU($n$), there will be an essentail singularity at $x=0$ or $x=\infty$, due to the wild behavior of $\theta_p(x)$ at the origin and the infinity, and the above method to derive the qq-characters will break down. This result agrees with that obtained from the discussion of gauge anomaly cancellation in \cite{Bhardwaj}. With a similar reasoning, Chern-Simons-like contributions are banned in the partition function. 

\paragraph{With $D$-type quiver?}

A reflection state constructed in \cite{D-type} can also be used in the elliptic DIM algebra to build gauge theories with $D$-type or affine $D$-type quiver. 6d theories with affine $D$-type quiver is classified in \cite{Bhardwaj} as one of the little string theories. It will be very interesting to investigate these theories in details with the elliptic vertex in the future.

There appeared different approaches to obtain the Nekrasov partition function by using the blowup equation \cite{blowup} for generic Calabi-Yau manifolds and the $(q,t)$-KZ equations \cite{ellipticKZ}. It would be extremely interesting to see the connection with these works.  

\section*{Acknowledgement}

The author would like to thank Yutaka Matsuo, Taro Kimura, Hironori Mori, Kimyeong Lee, Jean-Emile Bourgine, Omar Foda, Hong Zhang for inspiring discussions, and especially Yutaka Matsuo, Omar Foda for advice and comments on the draft. RZ thanks the mathematical research institute MATRIX in Australia where part of this research was performed at the occasion of the workshop ``Integrability in Low Dimensional Quantum Systems''. RZ is supported by JSPS fellowship for Young students, and thanks the hospitality of Yukawa Institute for Theoretical Physics during his short stay, when this project was initiated, and {\it East Asia Joint Workshop on Fields and Strings 2017 (KEK Theory Workshop 2017)} for giving him the opportunity to present part of the results in this work. 

\appendix 

\section{Trace over Fock Space}\label{a-trace}

We consider the correlation function in a thermalized harmonic oscillator, described by 
\ba
\lt[\alpha,\alpha^\dagger\rt]=1.
\ea
The modulus of the thermal circle is set to be $\tau$. 
Using the orthogonality of the Gaussian integral, 
\ba
\int{\rm d}z{\rm d}\bar{z}\ z^n\bar{z}^me^{-z\bar{z}}=n!\delta_{n,m},
\ea
where we chose the normalization of the measure to be ${\rm d}z{\rm d}\bar{z}=\frac{1}{\pi}r{\rm d}r{\rm d}\theta$, we can rewrite the thermalized correlation function for $:\phi(\alpha,\alpha^\dagger):$ to 
\ba
{\rm tr}_\tau:\phi(\alpha,\alpha^\dagger):=\sum_{n\geq 0}\frac{1}{n!}e^{-n\tau}\bra{0}\alpha^n:\phi(\alpha,\alpha^\dagger):\alpha^{\dagger n}\ket{0}\nn\\
=\int{\rm d}z{\rm d}\bar{z}\ e^{-z\bar{z}}\bra{0}e^{\alpha \bar{z}}:\phi(\alpha,\alpha^\dagger):e^{e^{-\tau}\alpha^\dagger z}\ket{0}.
\ea
Using the property of the coherent state 
\ba
\alpha e^{e^{-\tau}\alpha^\dagger z}\ket{0}=ze^{-\tau}e^{e^{-\tau}\alpha^\dagger z}\ket{0},\quad \bra{0}e^{\alpha\bar{z}}\alpha^\dagger=\bra{0}e^{\alpha\bar{z}} \bar{z},
\ea
the correlation funciton can be converted into a usual integral, 
\ba
{\rm tr}_\tau:\phi(\alpha,\alpha^\dagger):=\int{\rm d}z{\rm d}\bar{z}\ e^{-(1-e^{-\tau})z\bar{z}}\phi(e^{-\tau}z,\bar{z}).
\ea

A useful example is 
\ba
:\phi(\alpha,\alpha^\dagger):=\exp\lt(c\alpha\rt)\exp\lt(d\alpha^\dagger\rt),
\ea
for which 
\ba
{\rm tr}_\tau:\phi(\alpha,\alpha^\dagger):=\frac{1}{1-e^{-\tau}}\exp\lt(\frac{e^{-\tau}}{1-e^{-\tau}}cd\rt).
\ea

For the Heisenberg algebra, 
\ba
\lt[\hat{a}_n,\hat{a}_m\rt]=n\delta_{n+m,0},
\ea
the torus correlation function is defined with the propagator $L_0=\sum_{n>0}\hat{a}_{-n}\hat{a}_n$ instead of $\alpha^\dagger \alpha$. We therefore have 
\ba
{\rm tr}_\tau :\prod_{n>0}\phi_n(\hat{a}_n,\hat{a}_{-n}):=\prod_{n>0}\int{\rm d}z_n{\rm d}\bar{z}_n\ e^{-(1-e^{-n\tau})z_n\bar{z}_n}\phi_n(e^{-n\tau}z_n,n\bar{z}_n).
\ea
Again for the example, 
\ba
\phi_n(\hat{a}_n,\hat{a}_{-n})=:\exp\lt(c_n\hat{a}_{-n}\rt)\exp\lt(d_n\hat{a}_n\rt):,
\ea
we have 
\ba
{\rm tr}_\tau :\prod_{n>0}\phi_n(\hat{a}_n,\hat{a}_{-n}):=\exp\lt(\sum_{n>0}n\frac{e^{-n\tau}}{1-e^{-n\tau}}c_nd_n-\log(1-e^{-n\tau})\rt).
\ea

\section{Properties of $\theta$-function}\label{s:theta}

Our definition of the $\theta$-function is 
\ba
\theta_p(z)=(z;p)_\infty(pz^{-1};p)_\infty.
\ea
The Laurent expansion of $\theta_p(z)$ over $z$ can be obtained from the Jacobi triple identity, 
\ba
\prod_{i=1}^\infty (1-x^{2i})(1+x^{2i-1}y)(1+x^{2i-1}y^{-1})=\sum_{n=-\infty}^{\infty}x^{n^2}y^n,
\ea
by putting $xy=-z$, $x^2=p$, 
\ba
(p;p)_\infty (z;p)_\infty(z^{-1}p;p)_\infty=\sum_{n=-\infty}^\infty (-1)^np^{\frac{n^2-n}{2}}z^n.\label{expand-theta}
\ea
One can find the proof of this identity in various literatures, for example a physically motivated derivation in \cite{Ginsparg}, and we clearly see that it is a well-defined Taylor expansion about $p$. Similarly when we encounter $\theta_p(z)^{-1}$, we can expand $\frac{1}{(zp;p)_\infty}$ and $\frac{1}{(z^{-1}p;p)_\infty}$ about $p$, with a prefactor $\frac{1}{1-z}=\sum_{n\in\mathbb{Z}_{\geq0}}z^n$ multiplied. Collecting all these contributions, we can obtain a double expansion about both $p$ and $z$ for a general function of the form 
\ba
\frac{\prod_i\theta_p(a_iz)}{\prod_j\theta_p(b_jz)}.\nn
\ea

From the definition of the $\theta$-function, it is straightforward to check that 
\ba
&&\theta_p(z^{-1})=-z^{-1}\theta_p(z),\label{id-theta-1}\\
&&\theta_p(zp^n)=(-z)^{-n}p^{-\frac{n(n-1)}{2}}\theta_p(z).
\ea

\section{Rewriting the Nekrasov Factor}\label{Nekra}

The elliptic version of the Nekrasov factor is defined as \cite{M-strings,ellip-Virasoro} (see also \cite{KMS}) 
\ba
Z_{bfd}^{ellip}(v_1,\lambda;v_2,\nu|\mu)=\prod_{(i,j)\in\lambda}\theta_p\lt((v_1/v_2)\mu^{-1}q_1^{\nu^t_j-i}q_2^{-\lambda_i+j-1}\rt)\prod_{(i,j)\in\nu}\theta_p\lt((v_1/v_2)\mu^{-1}q_1^{-\lambda^t_j+i-1}q_2^{\nu_i-j}\rt).
\ea
Using the recursive formulae (C.5) and (C.6) derived in \cite{5dBMZ}, we have 
\ba
\frac{Z_{bfd}^{ellip}(v_1,\lambda+x;v_2,\nu|\mu)}{Z_{bfd}^{ellip}(v_1,\lambda;v_2,\nu|\mu)}=\frac{\prod_{y\in A(\nu)}\theta(\chi_x\chi_y^{-1}q_3^{-1})}{\prod_{y\in R(\nu)}\theta_p(\chi_x\chi_y^{-1})},\\
\frac{Z_{bfd}^{ellip}(v_1,\lambda-x;v_2,\nu|\mu)}{Z_{bfd}^{ellip}(v_1,\lambda;v_2,\nu|\mu)}=\frac{\prod_{y\in R(\nu)}\theta(\chi_x\chi_y^{-1})}{\prod_{y\in A(\nu)}\theta_p(\chi_x\chi_y^{-1}q_3^{-1})},\\
\frac{Z_{bfd}^{ellip}(v_1,\lambda;v_2,\nu+x|\mu)}{Z_{bfd}^{ellip}(v_1,\lambda;v_2,\nu|\mu)}=\frac{\prod_{y\in A(\lambda)}\theta(\chi_x\chi_y^{-1})}{\prod_{y\in R(\lambda)}\theta_p(\chi_x\chi_y^{-1}q_3^{-1})},\\
\frac{Z_{bfd}^{ellip}(v_1,\lambda;v_2,\nu-x|\mu)}{Z_{bfd}^{ellip}(v_1,\lambda;v_2,\nu|\mu)}=\frac{\prod_{y\in R(\lambda)}\theta_p(\chi_x\chi_y^{-1}q_3^{-1})}{\prod_{y\in A(\lambda)}\theta(\chi_x\chi_y^{-1})}.
\ea
Combining these identities with $Z^{ellip}_{vect}(v,\lambda)=Z^{ellip}_{bfd}(v,\lambda;v,\lambda|1)^{-1}$, we obtain 
\ba
\frac{Z^{ellip}_{vect}(v,\lambda+x)}{Z^{ellip}_{vect}(v,\lambda)}=\frac{1}{\theta_p(q_1)\theta_p(q_2)}\frac{\prod_{y\in R(\lambda)}\theta_p(\chi_x\chi_y^{-1})\theta_p(\chi_y\chi_x^{-1}q_3^{-1})}{\prod_{\substack{y\in A(\lambda)\\y\neq x}}\theta_p(\chi_x\chi_y^{-1}q_3^{-1})\theta_p(\chi_y\chi_x^{-1})},\\
\frac{Z^{ellip}_{vect}(v,\lambda-x)}{Z^{ellip}_{vect}(v,\lambda)}=\frac{1}{\theta_p(q_1)\theta_p(q_2)}\frac{\prod_{y\in A(\lambda)}\theta_p(\chi_x\chi_y^{-1}q_3^{-1})\theta_p(\chi_y\chi_x^{-1})}{\prod_{\substack{y\in R(\lambda)\\y\neq x}}\theta_p(\chi_x\chi_y^{-1})\theta_p(\chi_y\chi_x^{-1}q_3^{-1})}.
\ea
Starting from the initial value $Z_{bfd}^{ellip}(v_1,\emptyset;v_2,\emptyset|\mu)=1$, it is straight forward to derive the equivalent expression, which is repeatedly used in the Awata-Feigin-Shiraishi approach, 
\ba
Z_{bfd}^{ellip}(v_1,\lambda;v_2,\nu|\mu)=\prod_{x\in\lambda}\theta_p\lt(\frac{\chi_x\mu^{-1}}{v_2q_3}\rt)\prod_{y\in\nu}\theta_p\lt(\frac{v_1\mu^{-1}}{\chi_y}\rt)\prod_{x\in\lambda,y\in\nu}\frac{\theta_p(q\chi_x\chi_y^{-1}\mu^{-1})\theta_p(t^{-1}\chi_x\chi_y^{-1}\mu^{-1})}{\theta_p(\chi_x\chi_y^{-1}\mu^{-1})\theta_p(\chi_x\chi_y^{-1}q_3^{-1}\mu^{-1})}.\label{AFS-Nekra}
\ea
In the derivation of the above expression, we used the elliptic version of the shell formula, 
\ba
\frac{\prod_{x\in A(\lambda)}\theta_p(\chi_x/z)}{\prod_{x\in R(\lambda)}\theta_p(\chi_x/z)}=\theta_p(v/z)\prod_{x\in\lambda}\frac{\theta_p(q_1\chi_x/z)\theta_p(q_2\chi_x/z)}{\theta_p(\chi_x/z)\theta_p(q_3^{-1}\chi_x/z)},\label{shell-formula}
\ea
where $v=\chi_{(1,1)}$ is the Coulomb parameter associated to the Young diagram $\lambda$, and the identity $\theta_p(z^{-1})=-z^{-1}\theta_p(z)$.

Another well-known equivalent form of the Nekrasov factor is frequently used in the Iqbal-Kozcaz-Vafa approach \cite{IKV}, and its elliptic version is simply given by 
\ba
Z_{bfd}^{ellip}(v_1,\lambda;v_2,\nu|\mu)=\prod_{i,j=1}^\infty \frac{\theta_p(v_1\mu^{-1}/v_2 q_1^{-\lambda_i^t+j} q_2^{\lambda_j-i+1})}{\theta_p(v_1\mu^{-1}/v_2 q_1^{j} q_2^{-i+1})}.\label{IKV-Nekra}
\ea

\section{Coproduct of Elliptic DIM}\label{a-coproduct}

The standard Drinfeld coproduct is given by, 
\begin{align}
\begin{split}\label{AFS_coproduct}
&\Delta(x^+(z))=x^+(z)\otimes 1+\psi^-(\hat{\gamma}_{(1)}^{1/2}z)\otimes x^+(\hat{\gamma}_{(1)}z),\\
&\Delta(x^-(z))=x^-(\hat{\gamma}_{(2)} z)\otimes \psi^+(\hat{\gamma}_{(2)}^{1/2}z)+1\otimes x^-(z),\\
&\Delta(\psi^+(z))=\psi^+(\hat{\gamma}_{(2)}^{1/2}z)\otimes\psi^+(\hat{\gamma}_{(1)}^{-1/2}z),\\
&\Delta(\psi^-(z))=\psi^-(\hat{\gamma}_{(2)}^{-1/2}z)\otimes\psi^-(\hat{\gamma}_{(1)}^{1/2}z),
\end{split}
\end{align}
where $\hat{\gamma}_{(1)}=\hat{\gamma}\otimes 1$ and $\hat{\gamma}_{(2)}=1\otimes \hat{\gamma}$. Let us confirm it by explicit computation. 

\ba
\Delta(x^+(z))\Delta(x^+(w))=x^+(z)x^+(w)\otimes 1+x^+(z)\psi^-(\gammah_{(1)}^{1/2}w)\otimes x^+(\gammah_{(1)}w)+\psi^-(\gammah_{(1)}^{1/2}z) x^+(w)\otimes x^+(\gammah_{(1)}z)\nn\\
+\psi^-(\gammah_{(1)}^{1/2}z)\psi^-(\gammah_{(1)}^{1/2}w)\otimes x^+(\gammah_{(1)}z)x^+(\gammah_{(1)}w)\nn\\
=g_{ellip}(z/w)x^+(w)x^+(z)\otimes 1+g_{ellip}(\gammah_{(1)}^{-1/2}\gammah_{(1)}^{1/2}w/z)^{-1}\psi^-(\gammah_{(1)}^{1/2}w)x^+(z)\otimes x^+(\gammah_{(1)}w)\nn\\
+g_{ellip}(\gammah_{(1)}^{-1/2}\gammah_{(1)}^{1/2}z/w)x^+(w)\psi^-(\gammah_{(1)}^{1/2}z) \otimes x^+(\gammah_{(1)}z)\nn\\
+g_{ellip}(\gammah_{(1)}z/(\gammah_{(1)}w))\psi^-(\gammah_{(1)}^{1/2}w)\psi^-(\gammah_{(1)}^{1/2}z)\otimes x^+(\gammah_{(1)}w)x^+(\gammah_{(1)}z)\nn\\
=g_{ellip}(z/w)\Delta(x^+(w))\Delta(x^+(z)),\nn
\ea
where in the last line, we used $g_{ellip}(z^{-1})=g_{ellip}(z)^{-1}$. 
\ba
\Delta(x^-(z))\Delta(x^-(w))=x^-(\hat{\gamma}_{(2)} z)x^-(\hat{\gamma}_{(2)} w)\otimes \psi^+(\hat{\gamma}_{(2)}^{1/2}z)\psi^+(\hat{\gamma}_{(2)}^{1/2}w)+x^-(\hat{\gamma}_{(2)} z)\otimes \psi^+(\hat{\gamma}_{(2)}^{1/2}z)x^-(w)\nn\\
+x^-(\hat{\gamma}_{(2)} w)\otimes x^-(z)\psi^+(\hat{\gamma}_{(2)}^{1/2}w)+1\otimes x^-(z)x^-(w)\nn\\
=g(\gammah_{(2)}z/(\gammah_{(2)}w))^{-1}x^-(\hat{\gamma}_{(2)} w)x^-(\hat{\gamma}_{(2)} z)\otimes \psi^+(\hat{\gamma}_{(2)}^{1/2}w)\psi^+(\hat{\gamma}_{(2)}^{1/2}z)\nn\\
+g_{ellip}(\gammah_{(2)}^{-1/2}\gammah_{(2)}^{1/2}z/w)^{-1}x^-(\hat{\gamma}_{(2)} z)\otimes x^-(w)\psi^+(\hat{\gamma}_{(2)}^{1/2}z)\nn\\
+g_{ellip}(\gammah_{(2)}^{-1/2}\hat{\gamma}_{(2)}^{1/2}w/z)x^-(\hat{\gamma}_{(2)} w)\otimes \psi^+(\hat{\gamma}_{(2)}^{1/2}w)x^-(z)+g_{ellp}(z/w)^{-1}1\otimes x^-(w)x^-(z)\nn\\
=g_{ellip}(z/w)^{-1}\Delta(x^-(w))\Delta(x^-(z)).\nn
\ea
$g_{ellip}(z^{-1})=g_{ellip}(z)^{-1}$ is again used. 
\ba
&&\Delta(\psi^+(z))\Delta(\psi^-(w))=\psi^+(\hat{\gamma}_{(2)}^{1/2}z)\psi^-(\hat{\gamma}_{(2)}^{-1/2}w)\otimes\psi^+(\hat{\gamma}_{(1)}^{-1/2}z)\psi^-(\hat{\gamma}_{(1)}^{1/2}w)\nn\\
&&=\frac{g_{ellip}(\gammah_{(1)}\hat{\gamma}_{(2)}^{1/2}z/(\hat{\gamma}_{(2)}^{-1/2}w))}{g_{ellip}(\gammah_{(1)}^{-1}\hat{\gamma}_{(2)}^{1/2}z/(\hat{\gamma}_{(2)}^{-1/2}w))}\frac{g_{ellip}(\gammah_{(2)}\hat{\gamma}_{(1)}^{-1/2}z/(\hat{\gamma}_{(1)}^{1/2}w))}{g_{ellip}(\gammah_{(2)}^{-1}\hat{\gamma}_{(1)}^{-1/2}z/(\hat{\gamma}_{(1)}^{1/2}w))}\psi^-(\hat{\gamma}_{(2)}^{-1/2}w)\psi^+(\hat{\gamma}_{(2)}^{1/2}z)\otimes\psi^-(\hat{\gamma}_{(1)}^{1/2}w)\psi^+(\hat{\gamma}_{(1)}^{-1/2}z)\nn\\
&&=\frac{g_{ellip}(\gammah_{(1)}\hat{\gamma}_{(2)}z/w)}{g_{ellip}(\gammah_{(1)}^{-1}\hat{\gamma}_{(2)}^{-1}z/w)}\psi^-(\hat{\gamma}_{(2)}^{-1/2}w)\psi^+(\hat{\gamma}_{(2)}^{1/2}z)\otimes\psi^-(\hat{\gamma}_{(1)}^{1/2}w)\psi^+(\hat{\gamma}_{(1)}^{-1/2}z).\nn
\ea
\ba
\Delta(\psi^\pm(z))\Delta(x^+(w))=\psi^\pm(\hat{\gamma}_{(2)}^{\pm 1/2}z)x^+(w)\otimes\psi^\pm(\hat{\gamma}_{(1)}^{\mp1/2}z)+\psi^\pm(\hat{\gamma}_{(2)}^{\pm 1/2}z)\psi^-(\gammah_{(1)}^{1/2}w)\otimes\psi^\pm(\hat{\gamma}_{(1)}^{\mp1/2}z)x^+(\gammah_{(1)}w)\nn\\
=g_{ellip}(\gammah_{(1)}^{\pm 1/2} \hat{\gamma}_{(2)}^{\pm 1/2}z/w)x^+(w)\psi^\pm(\hat{\gamma}_{(2)}^{\pm 1/2}z)\otimes\psi^\pm(\hat{\gamma}_{(1)}^{\mp1/2}z)\nn\\
+\lt\{\begin{array}{c}
\frac{g_{ellip}(\gammah_{(1)}\gammah_{(1)}^{1/2}w/(\hat{\gamma}_{(2)}^{ 1/2}z))}{g_{ellip}(\gammah_{(1)}^{-1}\gammah_{(1)}^{1/2}w/(\hat{\gamma}_{(2)}^{ 1/2}z)}g_{ellip}(\gammah_{(2)}^{1/2}\hat{\gamma}_{(1)}^{-1/2}z/(\gammah_{(1)}w))\\
g_{ellip}(\gammah_{(2)}^{-1/2}\hat{\gamma}_{(1)}^{1/2}z/(\gammah_{(1)}w))\\
\end{array}
\rt.\psi^-(\gammah_{(1)}^{1/2}w)\psi^\pm(\hat{\gamma}_{(2)}^{\pm 1/2}z)\otimes x^+(\gammah_{(1)}w)\psi^\pm(\hat{\gamma}_{(1)}^{\mp1/2}z)\nn\\
=g_{ellip}(\gammah_{(1)}^{\pm1/2}\gammah_{(2)}^{\pm1/2}z/w)\Delta(x^+(w))\Delta(\psi^\pm(z)).\nn
\ea
\ba
\Delta(\psi^\pm(z))\Delta(x^-(w))=\psi^\pm(\hat{\gamma}_{(2)}^{\pm 1/2}z)x^-(\gammah_{(2)}w)\otimes\psi^\pm(\hat{\gamma}_{(1)}^{\mp1/2}z)\psi^+(\gammah^{1/2}_{(2)}w)+\psi^\pm(\hat{\gamma}_{(2)}^{\pm 1/2}z)\otimes\psi^\pm(\hat{\gamma}_{(1)}^{\mp1/2}z)x^-(w)\nn\\
=\lt\{\begin{array}{c}
g_{ellip}(\gammah_{(1)}^{-1/2}\hat{\gamma}_{(2)}^{1/2}z/(\gammah_{(2)}w))^{-1}\\
g_{ellip}(\gammah_{(1)}^{1/2}\hat{\gamma}_{(2)}^{- 1/2}z/(\gammah_{(2)}w))^{-1}\frac{g_{ellip}(\gammah^{-1}_{(2)}\hat{\gamma}_{(1)}^{1/2}z/(\gammah^{1/2}_{(2)}w))}{g_{ellip}(\gammah_{(2)}\hat{\gamma}_{(1)}^{1/2}z/(\gammah^{1/2}_{(2)}w))}\\
\end{array}
\rt.x^-(\gammah_{(2)}w)\psi^\pm(\hat{\gamma}_{(2)}^{\pm 1/2}z)\otimes\psi^+(\gammah^{1/2}_{(2)}w)\psi^\pm(\hat{\gamma}_{(1)}^{\mp1/2}z)\nn\\
+g_{ellip}(\gammah_{(2)}^{\mp 1/2}\hat{\gamma}_{(1)}^{\mp1/2}z/w)^{-1}\psi^\pm(\hat{\gamma}_{(2)}^{\pm 1/2}z)\otimes x^-(w)\psi^\pm(\hat{\gamma}_{(1)}^{\mp1/2}z)\nn\\
=g_{ellip}(\gammah_{(2)}^{\mp 1/2}\hat{\gamma}_{(1)}^{\mp1/2}z/w)^{-1}\Delta(x^-(w))\Delta(\psi^\pm(z)).\nn
\ea
Finally we have, 
\ba
\lt[\Delta(x^+(z)),\Delta(x^-(w))\rt]=\lt[x^+(z)\otimes 1+\psi^-(\hat{\gamma}_{(1)}^{1/2}z)\otimes x^+(\hat{\gamma}_{(1)}z),x^-(\hat{\gamma}_{(2)} w)\otimes \psi^+(\hat{\gamma}_{(2)}^{1/2}w)+1\otimes x^-(w)\rt]\nn\\
=\frac{\theta_p(q_1)\theta_p(q_2)}{(p;p)^2_\infty \theta_p(q_1q_2)}\lt(\delta(\gammah_{(1)}\gammah_{(2)}w/z)\psi^+(\gammah_{(1)}^{1/2}\gammah_{(2)}w)-\delta(\gammah_{(1)}^{-1}\gammah_{(2)}w/z)\psi^-(\gammah^{-1/2}_{(1)}\gammah_{(2)}w)\rt)\otimes \psi^+(\hat{\gamma}_{(2)}^{1/2}w)\nn\\
+\frac{\theta_p(q_1)\theta_p(q_2)}{(p;p)^2_\infty \theta_p(q_1q_2)}\psi^-(\hat{\gamma}_{(1)}^{1/2}z)\otimes\lt(\delta(\gammah_{(2)}w/(\gammah_{(1)}z))\psi^+(\gammah^{1/2}_{(2)}w)-\delta(\gammah_{(2)}^{-1}w/(\gammah_{(1)}z))\psi^-(\gammah^{-1/2}_{(2)}w)\rt)\nn\\
=\frac{\theta_p(q_1)\theta_p(q_2)}{(p;p)^2_\infty \theta_p(q_1q_2)}\lt(\delta(\gammah_{(1)}\gammah_{(2)}w/z)\Delta(\psi^+(\gammah_{(1)}^{1/2}\gammah_{(2)}^{1/2}w))-\delta(\gammah_{(1)}^{-1}\gammah_{(2)}^{-1}w/z)\Delta(\psi^-(\gammah_{(1)}^{-1/2}\gammah_{(2)}^{-1/2}w))\rt)\nn.
\ea
We note that $\psi^-(\hat{\gamma}_{(1)}^{1/2}z)\otimes x^+(\hat{\gamma}_{(1)}z)$ commutes with $x^-(\hat{\gamma}_{(2)} w)\otimes \psi^+(\hat{\gamma}_{(2)}^{1/2}w)$ following from $g_{ellip}(z^{-1})=g_{ellip}(z)^{-1}$. We remark that the standard Drinfeld coproduct structure holds as long as $g_{ellip}(z)^{-1}=g_{ellip}(z^{-1})$ is satisfied.

\section{Generalized Ramanujan's Identity}\label{Ramanujan}

The generalized Ramanujan identity is given by 
\ba
\prod_{i=1}^N\frac{1}{\theta_p(x_i/z)}\prod_{j=1}^M\theta_p(y_j/z)-\prod_{i=1}^N\frac{-z/x_i}{\theta_p(z/x_i)}\prod_{j=1}^M\lt(-\frac{\theta_p(z/y_j)}{z/y_j}\rt)=\sum_{i}\frac{\delta(x_i/z)}{(p;p)^2_\infty}\prod_{k\neq i}\frac{1}{\theta_p(x_k/x_i)}\prod_{j=1}^M\theta_p(y_j/x_i).\nn\\\label{Ramanujan-identity}
\ea
{\bf Proof:} We rewrite 
\ba
\prod_{i=1}^N\frac{1}{\theta_p(x_i/z)}\prod_{j=1}^M\theta_p(y_j/z)=\prod_{i=1}^N\frac{1}{(1-x_i/z)(px_i/z;p)_\infty(pz/x_i;p)_\infty}\prod_{j=1}^M(1-y_j/z)(py_j/z;p)_\infty(pz/y_j;p)_\infty\nn\\
=\frac{\prod_{j=1}^M(1-y_j/z)}{\prod_{i=1}^N(1-x_i/z)}\frac{\prod_{j=1}^M(py_j/z;p)_\infty(pz/y_j;p)_\infty}{\prod_{i=1}^N(px_i/z;p)_\infty(pz/x_i;p)_\infty},\nn
\ea
and 
\ba
\prod_{i=1}^N\frac{-z/x_i}{\theta(z/x_i)}\prod_{j=1}^M\lt(-\frac{\theta_p(z/y_j)}{z/y_j}\rt)=\prod_{i=1}^N\frac{-z/x_i}{(1-z/x_i)}\prod_{j=1}^M\frac{(1-z/y_j)}{-z/y_j}\frac{\prod_{j=1}^M(py_j/z;p)_\infty(pz/y_j;p)_\infty}{\prod_{i=1}^N(px_i/z;p)_\infty(pz/x_i;p)_\infty}.\nn
\ea
There is a common factor in the above two terms, and the remaining part reduces to the well-known identity of $\delta$-function in the non-elliptic case, 
\ba
\frac{\prod_i(1-a_iz)}{\prod_j(1-b_jz)}-z^{|\{i\}|-|\{j\}|}\frac{\prod_i(z^{-1}-a_i)}{\prod_j(z^{-1}-b_j)}=\sum_k\frac{\prod_i(1-a_i/b_k)}{\prod_{j\neq k}(1-b_j/b_k)}\delta(b_k z).\label{delta-fractional}
\ea

\section{Check of Vertical Representation}\label{vert-rep}

As done in \cite{BFHMZ}, all defining relations of the elliptic DIM but the commutation relation of $\lt[x^+,x^-\rt]$ can be satisfied with the key property of the $\cY$-function, 
\ba
\frac{\cY^{ellip}_{\lambda+x}(z)}{\cY^{ellip}_{\lambda}(z)}=\frac{\theta_p(q\chi_x/z)\theta_p(t^{-1}\chi_x/z)}{\theta_p(\chi_x/z)\theta_p(q_3^{-1}\chi_x/z)}=:S^{ellip}(\chi_x/z),
\ea
following from the shell formula (\ref{shell-formula}), since again 
\ba
g^{ellip}(z)=\frac{S^{ellip}(z)}{S^{ellip}(q_3z)}.
\ea
The remaining relation to check is 
\ba
\lt[x^+(z),x^-(w)\rt]=\frac{\theta_p(q)\theta_p(t^{-1})}{(p;p)^2_\infty\theta_p(q_3^{-1})}\delta(w/z)\lt(\psi^+(z)-\psi^-(z)\rt).
\ea
We have 
\ba
\lt[x^+(z),x^-(w)\rt]\ket{v,\lambda}\rangle=\frac{\theta_p(q)\theta_p(t^{-1})\gamma^{-1}}{(p;p)_\infty^4\theta_p(q_3^{-1})}\delta(z/w)\lt[\sum_{x\in A(\lambda)}\delta(\chi_x/z)\Res_{z\rightarrow \chi_x}{}^\theta\frac{\cY^{ellip}_\lambda(zq_3^{-1})}{z\cY^{ellip}_\lambda(z)}\rt.\nn\\
\lt.+\sum_{x\in R(\lambda)}\delta(\chi_x/z)\Res_{z\rightarrow \chi_x}{}^\theta\frac{\cY^{ellip}_\lambda(zq_3^{-1})}{z\cY^{ellip}_\lambda(z)}\rt]\ket{v,\lambda}\rangle,\nn
\ea
and we recognize the r.h.s. as the r.h.s. of the generalized Ramanujan's identity (\ref{Ramanujan-identity}) for $\Psi^{ellip}_\lambda(z)$ by recalling the definition of the $\cY$-function (\ref{def-Y}) and the elliptic residue (\ref{ellip-Res}). Therefore let us denote the equation after the use of (\ref{Ramanujan-identity}) as 
\ba
\lt[x^+(z),x^-(w)\rt]\ket{v,\lambda}\rangle=\frac{\theta_p(q)\theta_p(t^{-1})\gamma^{-1}}{(p;p)_\infty^2\theta_p(q_3^{-1})}\delta(z/w)\lt[\Psi^{ellip}_\lambda(z)|_{|z|<1}-\Psi^{ellip}_\lambda(z)|_{|z|>1}\rt]\ket{v,\lambda}\rangle,
\ea
where 
\ba
\Psi^{ellip}_\lambda(z)|_{|z|<1}=\prod_{x\in A(\lambda),y\in R(\lambda)}\frac{\theta_p(q_3\chi_x/z)\theta_p(q_3^{-1}\chi_y/z)}{\theta_p(\chi_y/z)\theta_p(\chi_x/z)},\\
\Psi^{ellip}_\lambda(z)|_{|z|<1}=q_3^{-1}\prod_{x\in A(\lambda),y\in R(\lambda)}\frac{\theta_p(q_3^{-1}z/\chi_x)\theta_p(q_3z/\chi_y)}{\theta_p(z/\chi_y)\theta_p(z/\chi_x)},
\ea
and we note that we have to expand the above expressions honestly with the double-expansion (about $p$ and $z$) expression of $\theta_p(z)$ and $\theta_p(z)^{-1}$ to obtain the representation of $\psi^\pm_k$ on $\ket{v,\lambda}\rangle$. See Appendix \ref{s:theta} for more details on this expansion. 

\section{Check of Horizontal Representation}\label{hor-rep}

From 
\ba
\eta(z)\eta(w)=\frac{\theta_p(w/z)\theta_p(q_3^{-1}w/z)}{\theta_p(qw/z)\theta_p(t^{-1}w/z)}:\eta(z)\eta(w):,\\
\eta(w)\eta(z)=\frac{\theta_p(z/w)\theta_p(q_3^{-1}z/w)}{\theta_p(qz/w)\theta_p(t^{-1}z/w)}:\eta(z)\eta(w):,
\ea
we obtain 
\ba
\eta(z)\eta(w)=\frac{\theta_p(qz/w)\theta_p(t^{-1}z/w)\theta_p(q_3z/w)}{\theta_p(q^{-1}z/w)\theta_p(tz/w)\theta_p(q_3^{-1}z/w)}\eta(w)\eta(z),
\ea
which reproduces the commutation relation, 
\ba
x^+(z)x^+(w)=\frac{\theta_p(qz/w)\theta_p(t^{-1}z/w)\theta_p(q_3z/w)}{\theta_p(q^{-1}z/w)\theta_p(tz/w)\theta_p(q_3^{-1}z/w)}x^+(w)x^+(z).
\ea
The contraction between two $\xi$'s is 
\ba
\xi(z)\xi(w)=\frac{\theta_p(z/w)\theta_p(z/wq_3^{-1})}{\theta_p(t^{-1}z/w)\theta_p(qz/w)}:\xi(z)\xi(w):,
\ea
which has the same commutation relation as $x^-(z)x^-(w)$. 
\ba
\eta(z)\varphi(w)=\frac{\theta_p(t\gamma^{-1/2}w/z)\theta_p(q^{-1}\gamma^{-1/2}w/z)\theta_p(q_3^{-1}\gamma^{-1/2}w/z)}{\theta_p(t^{-1}\gamma^{-1/2}w/z)\theta_p(q\gamma^{-1/2}w/z)\theta_p(q_3\gamma^{-1/2}w/z)}:\eta(z)\varphi(w):,
\ea
and 
\ba
\varphi(w)\eta(z)=:\varphi(w)\eta(z):,
\ea
are combined into 
\ba
\varphi(z)\eta(w)=\frac{\theta_p(t^{-1}\gamma^{-1/2}z/w)\theta_p(q\gamma^{-1/2}z/w)\theta_p(q_3\gamma^{-1/2}z/w)}{\theta_p(t\gamma^{-1/2}z/w)\theta_p(q^{-1}\gamma^{-1/2}z/w)\theta_p(q_3^{-1}\gamma^{-1/2}z/w)}\eta(w)\varphi(z).
\ea
It has the same commutation relation as that between $\psi^-(z)$ and $x^+(w)$. 
\ba
\eta(z)\bar{\varphi}(w)=:\eta(z)\bar{\varphi}(w):,
\ea
and 
\ba
\bar\varphi(w)\eta(z)=\frac{\theta_p(t\gamma^{-1/2}z/w)\theta_p(q^{-1}\gamma^{-1/2}z/w)\theta_p(q_3^{-1}\gamma^{-1/2}z/w)}{\theta_p(t^{-1}\gamma^{-1/2}z/w)\theta_p(q\gamma^{-1/2}z/w)\theta_p(q_3\gamma^{-1/2}z/w)}:\bar\varphi(w)\eta(z):.
\ea
lead to 
\ba
\bar{\varphi}(z)\eta(w)=\frac{\theta_p(t^{-1}\gamma^{1/2} z/w)\theta_p(q\gamma^{1/2} z/w)\theta_p(q_3\gamma^{1/2} z/w)}{\theta_p(t\gamma^{1/2} z/w)\theta_p(q^{-1}\gamma^{1/2} z/w)\theta_p(q_3^{-1}\gamma^{1/2} z/w)}\eta(w)\bar{\varphi}(z),
\ea
which is the same commutation relation as that between $\psi^+(z)$ and $x^+(w)$. The contraction for $\varphi$ (resp. $\bar\varphi$) with $\xi(z)$ can be obtained by exchanging numerators and denominators in the computation for $\eta(z)$ and shifting $z^n$ by $\gamma^{-|n|}$: 
\ba
\varphi(z)\xi(w)=\lt(\frac{\theta_p(t^{-1}\gamma^{1/2}z/w)\theta_p(q\gamma^{1/2}z/w)\theta_p(q_3\gamma^{1/2}z/w)}{\theta_p(t\gamma^{1/2}z/w)\theta_p(q^{-1}\gamma^{1/2}z/w)\theta_p(q_3^{-1}\gamma^{1/2}z/w)}\rt)^{-1}\xi(w)\varphi(z),\\
\bar{\varphi}(z)\xi(w)=\lt(\frac{\theta_p(t^{-1}\gamma^{-1/2} z/w)\theta_p(q\gamma^{-1/2} z/w)\theta_p(q_3\gamma^{-1/2} z/w)}{\theta_p(t\gamma^{-1/2} z/w)\theta_p(q^{-1}\gamma^{-1/2} z/w)\theta_p(q_3^{-1}\gamma^{-1/2} z/w)}\rt)^{-1}\xi(w)\bar{\varphi}(z),
\ea
which indeed reproduce the commutation relation between $x^-$ and $\psi^\mp$ respectively. The last commutation relation $[x^+,x^-]$ fixes the prefactor of $\psi^\pm$. 
\ba
&&\eta(z)\xi(w)=\frac{\theta_p(qw\gamma/z)\theta_p(t^{-1}w\gamma/z)}{\theta_p(w\gamma/z)\theta_p(q_3^{-1}w\gamma/z)}:\eta(z)\xi(w):,\\
&&\xi(w)\eta(z)=\frac{\theta_p(qz\gamma/w)\theta_p(t^{-1}z\gamma/w)}{\theta_p(z\gamma/w)\theta_p(q_3^{-1}z\gamma/w)}:\xi(w)\eta(z):,\\
&&:\eta(z)\xi(w):=\exp\lt(\sum_{n\neq0}\frac{1-t^{-n}}{1-p^{|n|}}p^{|n|}\frac{b_n}{n}(\gamma^{-|n|}w^n-z^n)\rt)\exp\lt(\sum_{n\neq0}\frac{1-t^n}{1-p^{|n|}}\frac{a_n}{n}(\gamma^{|n|}w^{-n}-z^{-n})\rt):.\nn\\
\ea
We note that the r.h.s. in the first two equations above are exactly the same, but expanded in different ways with respect to $z/w$, exactly as in the generalized Ramanujan's identity (\ref{Ramanujan-identity}). Using (\ref{Ramanujan-identity}), we have
\ba
&&\lt[\eta(z),\xi(w)\rt]=\delta(\gamma w/z)\frac{\theta_p(q)\theta_p(t^{-1})}{(p;p)^2_\infty\theta_p(q_3^{-1})}:\exp\lt(\sum_{n>0}\frac{1-t^{-n}}{1-p^n}p^n\frac{b_{n}}{n}(\gamma^{-n}-\gamma^n)w^n\rt)\nn\\
&&\times\exp\lt(\sum_{n>0}\frac{1-t^n}{1-p^n}\frac{a_n}{n}(\gamma^n-\gamma^{-n})w^{-n}\rt):+\delta(\gamma z/w)\frac{\theta_p(qq_3)\theta_p(t^{-1}q_3)}{(p;p)^2_\infty\theta_p(q_3)}\nn\\
&&:\exp\lt(-\sum_{n>0}\frac{1-t^n}{1-p^n}p^n\frac{b_{-n}}{n}(\gamma^{-n}-\gamma^n)w^{-n}\rt)
\exp\lt(-\sum_{n>0}\frac{1-t^{-n}}{1-p^n}\frac{a_{-n}}{n}(\gamma^n-\gamma^{-n})w^n\rt):\nn\\
&&=\frac{\theta_p(q)\theta_p(t^{-1})}{(p;p)^2_\infty\theta_p(q_3^{-1})}\lt(\delta(\gamma w/z)\bar\varphi(\gamma^{1/2}w)-\delta(\gamma z/w)\varphi(\gamma^{-1/2}w)\rt).
\ea
We thus see that indeed the horizontal representation 
\ba
x^+(z)\mapsto u\gamma^{n}z^{-n}\eta(z),\quad x^-(z)\mapsto u^{-1}\gamma^{-n}z^{n}\xi(z),\quad \psi^+(z)\mapsto \gamma^{-n}\bar\varphi(z),\quad \psi^-(z)\mapsto \gamma^n\varphi(z),
\ea
satisfies the defining relations of the elliptic DIM. 

\section{Proof of AFS property}\label{AFS-prop}

To check the AFS property for these vertices, we need the following results of contractions between vertex operators. 
\ba
&&\contraction{}{\eta(z)}{}{\Phi_\lambda[u,v]}\eta(z)\Phi_\lambda[u,v]=\frac{1}{\cY_\lambda(z)}:\eta(z)\Phi_\lambda[u,v]:,\\
&&\contraction{}{\Phi_\lambda[u,v]}{}{\eta(z)}\Phi_\lambda[u,v]\eta(z)=-\frac{vq_3}{z}\frac{1}{\cY_\lambda(zq_3^{-1})}:\Phi_\lambda[u,v]\eta(z):,\\
&&\contraction{}{\xi(z)}{}{\Phi_\lambda[u,v]}\xi(z)\Phi_\lambda[u,v]=\cY_\lambda(\gamma^{-1}z):\xi(z)\Phi_\lambda[u,v]:,\\
&&\contraction{}{\Phi_\lambda[u,v]}{}{\xi(z)}\Phi_\lambda[u,v]\xi(z)=-\gamma^{-1}\frac{z}{v}\cY_\lambda(z\gamma^{-1}):\Phi_\lambda[u,v]\xi(z):,\\
&&\contraction{}{\varphi(z)}{}{\Phi_\lambda[u,v]}\varphi(z)\Phi_\lambda[u,v]=:\varphi(z)\Phi_\lambda[u,v]:,\\
&&\contraction{}{\Phi_\lambda[u,v]}{}{\varphi(z)}\Phi_\lambda[u,v]\varphi(z)=\gamma^2\frac{\cY_\lambda(z\gamma^{-1/2})}{\cY_\lambda(z\gamma^{-1/2}q_3^{-1})}:\Phi_\lambda[u,v]\varphi(z):,\\
&&\contraction{}{\bar{\varphi}(z)}{}{\Phi_\lambda[u,v]}\bar{\varphi}(z)\Phi_\lambda[u,v]=\frac{\cY_\lambda(z\gamma^{1/2} q_3^{-1})}{\cY_\lambda(z\gamma^{1/2})}:\bar{\varphi}(z)\Phi_\lambda[u,v]:,\\
&&\contraction{}{\Phi_\lambda[u,v]}{}{\bar{\varphi}(z)}\Phi_\lambda[u,v]\bar{\varphi}(z)=:\Phi_\lambda[u,v]\bar{\varphi}(z):,
\ea
\ba
&&\contraction{}{\eta(z)}{}{\Phi^\ast_\lambda[u,v]}\eta(z)\Phi^\ast_\lambda[u,v]=\cY_\lambda(z\gamma^{-1}):\eta(z)\Phi^\ast_\lambda[u,v]:,\\
&&\contraction{}{\Phi^\ast_\lambda[u,v]}{}{\eta(z)}\Phi^\ast_\lambda[u,v]\eta(z)=-\gamma^{-1}z/v\cY_\lambda(z\gamma^{-1}):\Phi^\ast_\lambda[u,v]\eta(z):,\\
&&\contraction{}{\xi(z)}{}{\Phi^\ast_\lambda[u,v]}\xi(z)\Phi^\ast_\lambda[u,v]=\frac{1}{\cY_\lambda(zq_3^{-1})}:\xi(z)\Phi^\ast_\lambda[u,v]:,\\
&&\contraction{}{\Phi^\ast_\lambda[u,v]}{}{\xi(z)}\Phi^\ast_\lambda[u,v]\xi(z)=-\frac{v}{z}\frac{1}{\cY_\lambda(z)}:\Phi^\ast_\lambda[u,v]\xi(z):,\\
&&\contraction{}{\varphi(z)}{}{\Phi^\ast_\lambda[u,v]}\varphi(z)\Phi^\ast_\lambda[u,v]=:\varphi(z)\Phi^\ast_\lambda[u,v]:,\\
&&\contraction{}{\Phi^\ast_\lambda[u,v]}{}{\varphi(z)}\Phi^\ast_\lambda[u,v]\varphi(z)=\gamma^{-2}\frac{\cY_\lambda(z\gamma^{-3/2})}{\cY_\lambda(z\gamma^{1/2})}:\Phi^\ast_\lambda[u,v]\varphi(z):,\\
&&\contraction{}{\bar{\varphi}(z)}{}{\Phi^\ast_\lambda[u,v]}\bar{\varphi}(z)\Phi^\ast_\lambda[u,v]=\frac{\cY_\lambda(z\gamma^{-1/2})}{\cY_\lambda(z\gamma^{-5/2})}:\bar{\varphi}(z)\Phi^\ast_\lambda[u,v]:,\\
&&\contraction{}{\Phi^\ast_\lambda[u,v]}{}{\bar{\varphi}(z)}\Phi^\ast_\lambda[u,v]\bar{\varphi}(z)=:\Phi^\ast_\lambda[u,v]\bar{\varphi}(z):.
\ea

We check the AFS property for $g=x^\pm(z),\ \psi^\pm(z)$ one by one as follows. 
We have 
 \ba
 &&\rho^{(1,n+1)}_{-uv}(x^+(z))\Phi^{(n)}[u,v]-\Phi^{(n)}[u,v](\rho^{(0,1)}_{v}\otimes \rho^{(1,n)}_u)\lt(\psi^-(z)\otimes x^+(z)\rt)\nn\\
 &&=-\sum_{\lambda}a_\lambda uv\gamma^{n+1}z^{-(n+1)}\contraction{}{\eta(z)}{}{\Phi^{(n)}_\lambda}\eta(z)\Phi^{(n)}_\lambda \langle\bra{v,\lambda}-\sum_\lambda a_\lambda u\gamma^{n}z^{-n}\gamma^{-1}\Psi_\lambda\contraction{}{\Phi^{(n)}_\lambda}{}{\eta(z)}\Phi^{(n)}_\lambda\eta(z) \langle \bra{v,\lambda}\nn\\
 &&=-uv\gamma^{n+1}z^{-(n+1)}\sum_\lambda a_\lambda \lt(\lt.\frac{1}{\cY_\lambda(z)}\rt|_{z\sim\infty}-\lt.\frac{1}{\cY_\lambda(z)}\rt|_{z\sim0}\rt):\eta(z)\Phi^{(n)}_\lambda[u,v]: \langle \bra{v,\lambda},\nn
 \ea
 which exactly matches, 
 \ba
 &&\Phi^{(n)}[u,v]\rho^{(0,1)}_v(x^+(z))=\sum_\lambda a_\lambda \Phi^{(n)}_\lambda[u,v]\langle\bra{v,\lambda}x^+(z)\nn\\
 &&=-\frac{\gamma^{-1}b}{(p;p)^2_\infty}\sum_\lambda \sum_{x\in R(\lambda)}a_\lambda \Phi^{(n)}_\lambda \langle\bra{v,\lambda-x}\delta(z/\chi_x)\Res_{z\rightarrow \chi_x}{}^\theta z^{-1}\cY_\lambda(zq_3^{-1})\nn\\
 &&=-\frac{\gamma^{-1}b}{(p;p)^2_\infty}\sum_{\lambda'}\sum_{x\in A(\lambda')}a_{\lambda'}\frac{a_{\lambda'+x}}{a_{\lambda'}}\Phi^{(n)}_{\lambda'+x}\langle\bra{v,\lambda'}\delta(z/\chi_x)\frac{\theta_p(t)\theta_p(q^{-1})}{\theta_p(q_3)}\cY_{\lambda'}(\chi_xq_3^{-1})\nn\\
 &&=-uv\gamma^{n+1}\frac{b}{(p;p)^2_\infty}\sum_{\lambda'}\sum_{x\in A(\lambda')}a_{\lambda'}\chi_x^{-(n+1)}:\eta(\chi_x)\Phi^{(n)}_{\lambda'}:\langle\bra{v,\lambda'}\delta(z/\chi_x)\Res_{z\rightarrow \chi_x}{}^\theta\frac{1}{z\cY_{\lambda'}(z)},\nn
 \ea
 by using the generalized Ramanujan's identity (\ref{Ramanujan-identity}) and setting $b=1$. 
 
 For $x^-$, we have 
 \ba
&& \rho^{(1,n+1)}_{-uv}(x^-(z))\Phi^{(n)}[u,v]-\Phi^{(n)}[u,v](\rho^{(0,1)}_{v}\otimes \rho^{(1,n)}_u)\lt(1\otimes x^-(z)\rt)\nn\\
 &&=-\sum_{\lambda}a_\lambda u^{-1}v^{-1}\gamma^{-(n+1)}z^{n+1}\contraction{}{\xi(z)}{}{\Phi^{(n)}_\lambda}\xi(z)\Phi^{(n)}_\lambda \langle\bra{v,\lambda}-\sum_\lambda a_\lambda u^{-1}\gamma^{-n}z^{n}\contraction{}{\Phi^{(n)}_\lambda}{}{\xi(z)}\Phi^{(n)}_\lambda\xi(z) \langle \bra{v,\lambda}\nn\\
 &&=-\sum_\lambda a_\lambda u^{-1}v^{-1}\gamma^{-(n+1)}z^{n+1}\lt(\lt.\cY_\lambda(\gamma^{-1}z)\rt|_{z\sim\infty}-\lt.\cY_\lambda(\gamma^{-1}z)\rt|_{z\sim0}\rt):\xi(z)\Phi^{(n)}_\lambda[u,v]: \langle \bra{v,\lambda},\nn\\
 \ea
 matching with 
 \ba
 &&\Phi^{(n)}[u,v](\rho^{(0,1)}_{v}\otimes \rho^{(1,n)}_u)\lt(x^-(\gamma z)\otimes \psi^+(\gamma^{1/2}z)\rt)\nn\\
 &&=\frac{b^{-1}}{(p;p)^2_\infty}\gamma^{-n}\sum_\lambda\sum_{x\in A(\lambda)}a_\lambda:\Phi^{(n)}_\lambda[u,v]\bar{\varphi}(\gamma^{-1/2} \chi_x):\langle\bra{v,\lambda+x}\delta(\gamma z/\chi_x)\Res_{z\rightarrow \chi_x}{}^\theta \frac{1}{z\cY_\lambda(z)}\nn\\
 &&=-\frac{b^{-1}}{(p;p)^2_\infty}u^{-1}v^{-1}\gamma^{-(n+1)}\sum_{\lambda'}\sum_{x\in R(\lambda')}a_{\lambda'}(\chi_x/\gamma)^{n+1}:\Phi^{(n)}_{\lambda'}[u,v]\xi(\gamma^{-1} \chi_x):\langle\bra{v,\lambda'}\delta(\gamma z/\chi_x)\Res_{z\rightarrow \chi_x}{}^\theta z^{-1}\cY_\lambda(zq_3^{-1})\nn,
 \ea
 where we used 
 \ba
 \bar{\varphi}(z)=:\eta(\gamma^{1/2}z)\xi(\gamma^{-1/2}z):,
 \ea
 and 
 \ba
 \Res_{z\rightarrow \chi_x/\gamma}{}^\theta z^{-1}\cY_\lambda(z\gamma^{-1})= \Res_{z\rightarrow \chi_x}{}^\theta z^{-1}\cY_\lambda(zq_3^{-1}).\label{res-id}
 \ea
 
 For $\psi^\pm$, 
 \ba
 \rho^{(1,n+1)}_{-uv}(\psi^+(z))\Phi^{(n)}[u,v]=\sum_\lambda a_\lambda\gamma^{-(n+1)}\Psi_\lambda(z\gamma^{1/2}):\bar{\varphi}(z)\Phi^{(n)}_\lambda[u,v]:\langle\bra{v,\lambda},\\
 \rho^{(1,n+1)}_{-uv}(\psi^-(z))\Phi^{(n)}[u,v]=\sum_\lambda a_\lambda\gamma^{n+1}:\varphi(z)\Phi^{(n)}_\lambda[u,v]:\langle\bra{v,\lambda},
 \ea
 and 
 \ba
 &&\Phi^{(n)}[u,v](\rho^{(0,1)}_{v}\otimes \rho^{(1,n)}_u)\lt(\psi^+(\gamma^{1/2}z)\otimes \psi^+(z)\rt)\nn\\
 &&=\sum_\lambda a_\lambda \gamma^{-n}\gamma^{-1}\Psi_\lambda(z\gamma^{1/2}):\Phi^{(n)}_\lambda[u,v]\bar{\varphi}(z):\langle\bra{v,\lambda},\\
 &&\Phi^{(n)}[u,v](\rho^{(0,1)}_{v}\otimes \rho^{(1,n)}_u)\lt(\psi^-(\gamma^{-1/2}z)\otimes \psi^-(z)\rt)\nn\\
 &&=\sum_\lambda a_\lambda\gamma^n\gamma^{-1}\Psi_\lambda(z\gamma^{-1/2})\gamma^2\Psi^{-1}_\lambda(z\gamma^{-1/2}:\Phi^{(n)}_\lambda[u,v]\varphi(z):\langle\bra{v,\lambda}.
 \ea
 They agree with each other again, and we see that the AFS property holds for $\Phi^{(n)}[u,v]$. 
 
 For $\Phi^\ast$, 
 The confirmation for $g=x^+$ goes as, 
 \ba
  &&(\rho^{(1,n)}_{u}\otimes \rho^{(0,1)}_{v})(x^+(z)\otimes 1)\Phi^{\ast(n)}[u,v]-\Phi^{\ast(n)}[u,v]\rho^{(1,n+1)}_{-uv}(x^+(z))\nn\\
  &&=\sum_\lambda a_\lambda\ket{v,\lambda}\rangle\lt(u\gamma^{n}z^{-n}\lt.\cY_\lambda(z\gamma^{-1})\rt|_{z\sim\infty}:\eta(z)\Phi^{\ast(n)}_\lambda[u,v]:\rt.\nn\\
  &&\lt.+uv\gamma^{n+1}z^{-(n+1)}(-\gamma^{-1}z/v)\lt.\cY_\lambda(z\gamma^{-1})\rt|_{z\sim 0}:\Phi^{\ast(n)}_\lambda[u,v]\eta(z):\rt)\nn\\
  &&=\sum_\lambda a_\lambda \ket{v,\lambda}\rangle u\gamma^nz^{-n}\lt(\lt.\cY_\lambda(z\gamma^{-1})\rt|_{z\sim\infty}-\lt.\cY_\lambda(z\gamma^{-1})\rt|_{z\sim 0}\rt):\eta(z)\Phi^{\ast(n)}_\lambda[u,v]:,
  \ea
  and
  \ba
  &&(\rho^{(1,n)}_{u}\otimes \rho^{(0,1)}_{v})(\psi^-(\gamma^{1/2}z)\otimes x^+(\gamma z))\Phi^{\ast(n)}[u,v]\nn\\
  &&=\frac{\gamma^nb}{(p;p)^2_\infty}\sum_\lambda a_\lambda \sum_{x\in A(\lambda)}\delta(z\gamma/\chi_x)\Res_{z\rightarrow\chi_x}{}^\theta \frac{1}{z\cY_\lambda(z)}\ket{v,\lambda+x}\rangle:\varphi(z\gamma^{1/2})\Phi^{\ast(n)}_\lambda[u,v]:\nn\\
  &&=-\frac{\gamma^nb}{(p;p)^2_\infty}\sum_{\lambda'}\sum_{x\in R(\lambda')}a_{\lambda'}\gamma^{-1}\chi_x^{-1}\delta(z\gamma/\chi_x)\Res_{z\rightarrow\chi_x}{}^\theta \cY_{\lambda'}(zq_3^{-1})\ket{v,\lambda'}\rangle\gamma u(\gamma/\chi_x)^n:\eta(\chi_x\gamma^{-1})\Phi^{\ast(n)}_{\lambda'}[u,v]:\nn\\
  &&=-\frac{u\gamma^nb}{(p;p)^2_\infty}\sum_{\lambda'}\sum_{x\in R(\lambda')}a_{\lambda'}\delta(z\gamma/\chi_x)z^{-n}\Res_{z\rightarrow\chi_x}{}^\theta z^{-1}\cY_{\lambda'}(zq_3^{-1})\ket{v,\lambda'}\rangle:\eta(z)\Phi^{\ast(n)}_{\lambda'}[u,v]:,
  \ea
  where we used 
  \ba
  \varphi(z)=:\eta(z\gamma^{-1/2})\xi(z\gamma^{1/2}),
  \ea
  and (\ref{res-id}) again. 
  
For $g=x^-$, we have 
\ba
&&(\rho^{(1,n)}_{u}\otimes \rho^{(0,1)}_{v})(x^-(z)\otimes \psi^+(z))\Phi^{\ast(n)}[u,v]-\Phi^{\ast(n)}[u,v]\rho^{(1,n+1)}_{-uv}(x^-(z))\nn\\
&&=u^{-1}\gamma^{-n}z^n\sum_\lambda a_\lambda \gamma^{-1}\Psi_\lambda(z)\ket{v,\lambda}\rangle \frac{1}{\cY_\lambda(zq_3^{-1})}:\xi(z)\Phi^{\ast(n)}_\lambda[u,v]:\nn\\
&&-u^{-1}v^{-1}\gamma^{-(n+1)}z^{n+1}\sum_\lambda a_\lambda\ket{v,\lambda}\rangle\frac{v}{z}\frac{1}{\cY_\lambda(z)}:\xi(z)\Phi^{\ast(n)}_\lambda[u,v]: \nn\\
&&=u^{-1}\gamma^{-(n+1)}z^n\sum_\lambda a_\lambda \lt(\lt.\frac{1}{\cY_\lambda(z)}\rt|_{z\sim\infty}-\lt.\frac{1}{\cY_\lambda(z)}\rt|_{z\sim0}\rt)\ket{v,\lambda}\rangle:\xi(z)\Phi^{\ast(n)}_\lambda[u,v]: ,
\ea
matching with 
\ba
&&(\rho^{(1,n)}_{u}\otimes \rho^{(0,1)}_{v})(1\otimes x^-(z))\Phi^{\ast(n)}[u,v]\nn\\
&&=\frac{b^{-1}\gamma^{-1}}{(p;p)^2_\infty}\sum_\lambda a_\lambda\sum_{x\in R(\lambda)}\delta(z/\chi_x)\Res_{z\rightarrow \chi_x}{}^\theta z^{-1}\cY_\lambda(zq_3^{-1})\ket{v,\lambda-x}\rangle \Phi^{\ast(n)}_\lambda[u,v]\nn\\
&&=-\frac{b^{-1}\gamma^{-1}}{(p;p)^2_\infty}\sum_{\lambda'}\sum_{x\in A(\lambda')}a_{\lambda'}\delta(z/\chi_x)\gamma \chi_x^{-1}\Res_{z\rightarrow \chi_x}{}^\theta\frac{1}{\cY_{\lambda'}(z)}\ket{v,\lambda'}\rangle \gamma^{-1} u^{-1}\chi_x^n/\gamma^n:\xi(\chi_x)\Phi^{\ast(n)}_{\lambda'}[u,v]:\nn\\
&&=-\frac{b^{-1}\gamma^{-(n+1)}}{(p;p)^2_\infty}\sum_{\lambda'}\sum_{x\in A(\lambda')}a_{\lambda'}\delta(z/\chi_x)z^n\Res_{z\rightarrow \chi_x}{}^\theta\frac{1}{z\cY_{\lambda'}(z)}\ket{v,\lambda'}\rangle:\xi(\chi_x)\Phi^{\ast(n)}_{\lambda'}[u,v]:.
\ea

For $g=\psi^\pm$, we have 
\ba
&&(\rho^{(1,n)}_{u}\otimes \rho^{(0,1)}_{v})(\psi^+(z)\otimes \psi^+(\gamma^{-1/2}z))\Phi^{\ast(n)}[u,v]\nn\\
&&=\sum_\lambda a_\lambda \gamma^{-1}\Psi_\lambda(z\gamma^{-1/2})\ket{v,\lambda}\rangle \gamma^{-n}\frac{\cY_\lambda(z\gamma^{-1/2})}{\cY_\lambda(zq_3^{-1}\gamma^{-1/2}}:\bar{\varphi}(z)\Phi^{\ast(n)}_\lambda[u,v]:,\\
&&(\rho^{(1,n)}_{u}\otimes \rho^{(0,1)}_{v})(\psi^-(z)\otimes \psi^-(\gamma^{1/2}z))\Phi^{\ast(n)}[u,v]\nn\\
&&=\sum_\lambda a_\lambda \gamma^{-1}\Psi_\lambda(z\gamma^{1/2})\ket{v,\lambda}\rangle\gamma^n:\varphi(z)\Phi^{\ast(n)}_\lambda[u,v]:,
\ea
respectively equal to 
\ba
\Phi^{\ast(n)}[u,v]\rho^{(1,n+1)}_{-uv}(\psi^+(z))=\gamma^{-(n+1)}\sum_\lambda a_\lambda \ket{v,\lambda}\rangle :\Phi^{\ast(n)}_\lambda[u,v]\bar{\varphi}(z):,\\
\Phi^{\ast(n)}[u,v]\rho^{(1,n+1)}_{-uv}(\psi^+(z))=\gamma^{n+1}\sum_\lambda a_\lambda \ket{v,\lambda}\rangle \gamma^{-2}\Psi_\lambda(z\gamma^{1/2}):\Phi^{\ast(n)}_\lambda[u,v]\varphi(z):.
\ea
\begin{flushright}
Q.E.D.
\end{flushright}

\bibliography{ellipticvertex}

\end{document}